\documentclass[useAMS,usenatbib,aas_macros,fleqn]{mn2e}

\usepackage{epsfig}
\usepackage{amssymb}
\usepackage{natbib}
\usepackage{aas_macros}
\usepackage{times}

\usepackage{amsmath}

\DeclareGraphicsExtensions{.eps}

\newcommand{\figu}{Figure~}
\newcommand{\figus}{Figures~}
\newcommand{\eq}{Equation~}

\newcommand{\sect}{Section~}

\newcommand{\BH}{black hole}
\newcommand{\BHs}{black holes}

\newcommand{\ergse}{{\rm erg\, s^{-1}}}

\newcommand{\sis}{$\sigma$}
\newcommand{\sise}{\sigma}

\newcommand{\mbh}{$M_{\rm bh}$}
\newcommand{\mbhe}{M_{\rm bh}}

\newcommand{\mstar}{$M_{\rm star}$}
\newcommand{\mstare}{M_{\rm star}}
\newcommand{\mbulge}{$M_{\rm bulge}$}
\newcommand{\mbulgee}{M_{\rm bulge}}

\newcommand{\msune}{M_{\odot}}

\newcommand{\cmodel}{\textsc{cmodel}}
\newcommand{\SerExp}{\textsc{SerExp}}

\newcommand{\modI}{Model I}
\newcommand{\modII}{Model II}
\newcommand{\modIII}{Model III}

\begin{document}

\def\sarc{$^{\prime\prime}\!\!.$}
\def\arcsec{$^{\prime\prime}$}
\def\arcmin{$^{\prime}$}
\def\degr{$^{\circ}$}
\def\seco{$^{\rm s}\!\!.$}
\def\ls{\lower 2pt \hbox{$\;\scriptscriptstyle \buildrel<\over\sim\;$}}
\def\gs{\lower 2pt \hbox{$\;\scriptscriptstyle \buildrel>\over\sim\;$}}

\title[SMBHs: selection bias and its consequences]{Selection bias in dynamically-measured super-massive black hole samples: its consequences and the quest for the most fundamental relation}

\author[F. Shankar et al.]
{Francesco Shankar$^{1}$\thanks{E-mail:$\;$F.Shankar@soton.ac.uk},
Mariangela Bernardi$^{2}$, Ravi K. Sheth$^{2}$, Laura Ferrarese$^{3}$,
\newauthor
Alister W. Graham$^{4}$, Giulia Savorgnan$^{4}$, Viola Allevato$^{5}$, Alessandro Marconi$^{6}$,
\newauthor
Ronald L\"{a}sker $^{7}$, Andrea Lapi$^{8,9}$
\\
$1$ Department of Physics and Astronomy, University of Southampton, Highfield, SO17 1BJ, UK\\
$2$ Department of Physics and Astronomy, University of Pennsylvania, 209 South 33rd St, Philadelphia, PA 19104\\
$3$ Herzberg Institute of Astrophysics, National Research Council of Canada, Victoria, BC V9E 2E7, Canada\\
$4$ Centre for Astrophysics and Supercomputing, Swinburne University of Technology, Hawthorn, Victoria 3122, Australia\\
$5$ Department of Physics, University of Helsinki, Gustaf H\"{a}llstr\"{o}min katu 2a, FI-00014 Helsinki, Finland\\
$6$ Università degli Studi di Firenze, Dipartimento di Fisica e Astronomia, via G. Sansone 1, 50019 Sesto F.no, Firenze, Italy\\
$7$ Max Planck Institute for Astronomy, K\"{o}nigstuhl 17, D-69117 Heidelberg, Germany\\
$8$ SISSA, Via Bonomea 265, 34136 Trieste, Italy\\
$9$ Dip. Fisica, Univ. ``Tor Vergata'', Via Ricerca Scientifica 1, 00133 Roma, Italy}

\date{}
\pagerange{\pageref{firstpage}--
\pageref{lastpage}} \pubyear{2016}
\maketitle
\label{firstpage}

\begin{abstract}
We compare the set of local galaxies having dynamically measured \BHs\ with a large, unbiased sample of galaxies extracted from the Sloan Digital Sky Survey. We confirm earlier work showing that the majority of \BH\ hosts have significantly higher velocity dispersions \sis\ than local galaxies of similar stellar mass. We use Monte-Carlo simulations to illustrate the effect on \BH\ scaling relations if this bias arises from the requirement that the black hole sphere of influence must be resolved to measure black hole masses with spatially resolved kinematics.  We find that this selection effect artificially increases the normalization of the \mbh-\sis\ relation by a factor of at least $\sim 3$; the bias for the \mbh-\mstar\ relation is even larger. Our Monte Carlo simulations and analysis of the residuals from scaling relations both indicate that \sis\ is more fundamental than \mstar\ or effective radius.  In particular, the \mbh-\mstar\ relation is mostly a consequence of the \mbh-\sis\ and \sis-\mstar\ relations, and is heavily biased by up to a factor of 50 at small masses.  This helps resolve the discrepancy between dynamically-based black hole-galaxy scaling relations versus those of active galaxies. Our simulations also disfavour broad distributions of black hole masses at fixed \sis. Correcting for this bias suggests that the calibration factor used to estimate black hole masses in active galaxies should be reduced to values of $f_{vir}\sim1$.  Black hole mass densities should also be proportionally smaller, perhaps implying significantly higher radiative efficiencies/black hole spins. Reducing black hole masses also reduces the gravitational wave signal expected from black hole mergers.
\end{abstract}


\section{Introduction}
\label{sec|intro}
The presence of nuclear supermassive black holes (hereafter \BH) in the majority of local galaxies has become an accepted paradigm. Indeed, nuclear kinematics of a number of nearby galaxies show the clear signature of a central mass concentration, beyond what can be attributed to the observed stellar population in the nuclear regions.  Black hole masses, \mbh, are found to correlate with several global properties of their host galaxies \citep[see, e.g.,][for reviews]{FerrareseFord,ShankarReview,KormendyHo,GrahamReview15}, including the stellar and/or bulge mass, velocity dispersion, \sis, luminosity, light concentration or S\'{e}rsic index \citep[e.g.,][]{Magorrian98,Richstone98,Ferrarese00,Gebhardt00,Graham01,GrahamDriver,Marconi03,HaringRix,Graham07,Satyapal08,Graham12,KormendyHo,McConn13,Scott13,Laesker14,SavorgnanGraham,Savo15,Saglia16}, and the mass of the surrounding dark matter halo \citep[e.g.,][]{Ferrarese02,Baes03,Bogdan15,Sabra15}. However, while it is true that the number of dynamical \BH\ mass measurements has increased over the years, such samples still remain relatively small, of the order of $\sim 70-80$ galaxies.  This is due primarily to the difficulty of carrying out direct measurements with the required depth and spatial resolution \citep[see,  e.g.,][for reviews on the challenges encountered in these observational campaigns]{Faber99,FerrareseFord}.

Understanding the origin and reliability of these correlations is vital if we want to ultimately improve our understanding of galaxy-\BH\ (co-)evolution. For instance, the normalization and slope of the \mbh\--\sis\ relation may contain key information on whether feedback is primarily via energy or momentum transfer \citep[e.g.,][]{SilkRees,Fabian99,King05,WL05,Fabian12,King14}. Directly related to the normalization of the black hole scaling relations is the value of the virial $f_{\rm vir}$-factor used to derive the masses of black holes probed via reverberation mapping studies \citep[][]{Onken04,Veste06}.  If the \mbh\--\sis\ normalization is too high by some amount, then the $f_{\rm vir}$-factor will be too high by this same amount.  Shifting the normalization to lower masses will not only lower the quasar masses inferred at high-$z$, thus helping to solve the problem of the time required to grow the black hole by accretion, but an abundance of lower-mass black holes, now with $\mbhe < 10^5\ \msune$ (e.g, ``intermediate mass black holes''), will be realized, joining the ranks of objects like HLX-1 in ESO 243-49 \citep[][]{Farrell09,Farrell14,Webb14} and NGC~2276-3c \citep[][]{Mezcua15}.

The normalization of the \mbh-\mbulge\ relation is also a key ingredient in predicting what pulsar timing array searches \citep[e.g.,][]{Sesana08,Hobbs10,Sesana13b,Kramer13,Rosado14,Rosado15} for gravitational radiation will see \citep[][]{Bonnor61,Peres62,Beke73,Buonanno00,Berti09}.  With the normalizations currently in use, the pulsar timing arrays were expected to have detected a gravitational wave background \citep[][]{Shannon13,Shannon15}.  To explain the lack of detection, theorists have begun to consider new possibilities, like rather eccentric orbits for the coalescing binary supermassive black hole population so as to shift the gravitational wave spectral energy distribution out of the observing window of pulsar timing arrays.  However, eccentric orbits are at odds with the observed ellipticities of partially-depleted cores \citep[see, e.g.,][]{Dullo15}.  Environmental effects are also being invoked to reduce the time over which the binary emits gravitational radiation, and thus possibly resolve the dilemma \citep[e.g.,][]{Ravi14}.  However, if the \mbh-\mbulge\ normalization on which these arguments are based is too high, then the expected gravitational wave signal has been over-estimated.

Scatter in the black-hole galaxy scaling relations is thought to bear imprints of the amount of collisionless ``dry mergers'' experienced by the (most massive) hosts \citep[e.g.,][but see \citealt{SavorgnanGraham}]{Boylan06,Peng07,Hirschmann10,JahnkeMaccio}. The S\'{e}rsic index and the presence of a partially depleted core, along with their possible correlations with the mass of the central black hole, are also believed to contain information about the types of mergers responsible for shaping the host spheroids \citep[e.g.,][]{Aguerri01,Merritt06,Hilz13,Graham13,GrahamScott15}.

Beyond the local universe, data tracking the evolution of active and star-forming galaxies over cosmic time shows that \BH\ accretion and star formation peak at similar epochs \citep[e.g.,][]{Marconi04,Merloni04,Silverman08LF,Zheng09,SWM,delvecchio14}, consistent with the idea that massive \BHs\ and their host galaxies may be co-evolving.  One way to test this co-evolution is by exploring the cosmic evolution of the above scaling relations.  Thus, the characterization of the scaling relations of \BHs\ with their hosts is the subject of intense observational efforts, both locally and at high redshift \citep[e.g.,][]{Shields06,Lauer07bias,Treu07,Woo08,Gaskell09,ShankarMsigma,Merloni10,Schulze11,Falomo14,Shen15}.

In-depth knowledge of the \BH-host scaling relations at any epoch can also potentially provide statistical clues on the mass densities of \BHs.  For instance, a robust estimate of the \BH\ mass function can provide valuable constraints on the mechanisms governing \BH\ growth over cosmic time, such as mergers or disc instabilities \citep[e.g.,][]{KH00,Vittorini05,Bower06,Fontanot06,Lapi06,Menci06,Malbon07,SWM,Bournaud11b,Fanidakis11,Hirschmann12,Dubois13,Hirschmann14,Sesana14,Aversa15,FF15,Sija15}, as well as on the average radiative efficiencies/black hole spin and/or fraction of obscured sources \citep[e.g.,][]{Soltan,Elvis02,Shankar13acc,Aversa15,TucciVolo}.  However, because direct dynamical measurements of black hole masses are difficult to obtain, considerable effort has been invested in identifying easily observed proxies for \mbh. As an example, the standard procedure for calculating the \BH\ ``mass function'' has been to assume that all galaxies host \BHs, and to use the correlation between the observable proxy and \mbh\ to transform the observed distribution of the proxy into a distribution of \mbh\ \citep[e.g.,][]{Salucci99,AllerRichstone02,Ferrarese02demo,MD04,Marconi04,Shankar04,Benson07,Tundo07,Graham07BHMF,YuLu08,Vika09}.

This procedure rests on the assumption that one has correctly identified the observable proxy for \mbh, and that the scaling relation used to convert from it to \mbh\ has been correctly estimated.  For example, the two most commonly used proxies, stellar velocity dispersion and bulge luminosity, lead to rather different estimates of $\phi(\mbhe)$ \citep[e.g.,][but see also \citealt{Graham08FP}]{Lauer07demo,Tundo07}:  the luminosity-based estimate predicts many more massive black holes.  To date, there is no consensus on which is correct, at least for the more massive galaxies.  There is also no consensus on whether or not the best proxy for \mbh\ involves more than one observable. For example, some groups \citep[e.g.][]{FeoliMele,Hop07FP} argue that $\mbhe \propto R^{2-\beta/2}\sigma^\beta$, with $R$ any characteristic (e.g., half-light) radius of the host galaxy and $\beta\approx 3$, whereas, on the basis of more recent samples, \cite{Beifiori12} report no compelling evidence for anything other than $\mbhe\propto \sigma^\beta$ with $\beta\approx 4$, in line with \citet{Graham08FP}.  Notice that the second parameter in the \citet{FeoliMele} formulation becomes less important when $\beta\to 4$.

However, the two issues above are coupled: one cannot properly address the observable proxy question if the scaling relations have been incorrectly estimated. While there exists a wealth of literature on measuring these relations in the local dynamical \BH\ samples \citep[e.g.,][]{Ford98,YT02,Novak06,Lauer07bias,Graham07,Bat10,Merloni10,Graham11,Beifiori12,Graham13,SW14}, the extent to which selection effects can bias these estimates has not been fully addressed.  This matters because, as pointed out by \citet{Bernardi07} almost a decade ago, the available \BH\ samples are not a representative subset of early-type galaxies:  their host galaxies have larger than expected velocity dispersions than early type galaxies of the same luminosity or stellar mass.  Although \cite{YT02} had also noted that $\sigma$-$L$ in \BH\ samples appeared to be biased -- and \citet{Remco15} have recently reconfirmed that \BH\ hosts tend to be the densest galaxies given their luminosity -- they ignored the implications for \BH\ scaling relations.  \citet{Bernardi07} used analytic arguments and Monte-Carlo simulations to show that this is unwise -- selection effects can heavily bias \BH\ scaling relations.

There is at least one obvious selection effect:  direct black hole mass estimates depend on resolving (at least approximately) the sphere of influence $r_{\rm infl}\equiv G\mbhe/\sigma^2$ of the black hole \citep[e.g.][]{Peebles72,Ford98,MerrittFerrareseProceedings,Barth04,Bat10,Graham11,Gultekin11}.  This, at fixed signal-to-noise ratio, becomes more difficult as the distance to the black hole increases. The first part of this paper is devoted to a study of this selection effect.  In the second, we address the question of which scaling relation is more fundamental.

When cosmological parameters are necessary, we set
$h=0.7$, $\Omega_m=0.3$, $\Omega_{\Lambda}=0.7$.

\section{Data}
\label{sec|data}

The galaxy sample used as the reference data in this study is the one collected and studied in \citet[][]{Meert15}, and we refer to that paper for full details.  Briefly, galaxies are selected from the Sloan Digital Sky Survey (SDSS) DR7 spectroscopic sample \citep{2009ApJS..182..543A} in the redshift range $0.05<z<0.2$, and with a morphology classification based on the Bayesian automated morphological classifier by \citet{Huertas11}.  The latter statistically quantifies the morphological appearance of a galaxy with probabilities $p$(E--S0) of being an elliptical (E), a lenticular (S0), and a spiral, based on several different criteria.  Unless otherwise noted, we will always define the sample of ellipticals/lenticulars as those SDSS galaxies with a $p$(E--S0) $>0.80$, though the exact cut chosen to select early-type galaxies in SDSS does not impact any of our conclusions.

Galaxy mass-to-light ratios are linear functions of colour (following \citealt{Bell03SEDs}; see \eq6 in \citealt{Bernardi10}), derived through Spectral Energy Distribution fitting using the \citet{BC03} synthesis population models, and converted to a \citet{Chabrier03} Initial Mass Function (IMF).  Stellar masses are obtained by multiplying these mass-to-light ratios by the luminosity.  \citet{Bernardi13,Bernardi12,Bernardi16} have emphasized that the choice of luminosity matters as much as the choice of IMF.  They provide three different estimates for the stellar masses: one based on the SDSS \cmodel $\,$ magnitude; another, based on fitting a single S\'{e}rsic profile \cite{Sersic63}; and a third, \SerExp, based on a combination of S\'{e}rsic and exponential light profiles.  Unless we specify otherwise, all (circularized) galaxy effective radii and luminosities -- and hence stellar masses -- which follow are based on their \SerExp\ fits (also see \citealt{Meert15}).  While this choice matters quantitatively, it makes no qualitative difference to our findings. In addition, each of the groups we discuss below uses a different way of estimating $\mstare$ (assumptions about star formation history, dust, etc...).  In principle, we should correct all to a common reference point.  However, once corrected to the same IMF, systematic biases in stellar masses are of order $\sim 0.1$~dex \cite{Bernardi16ML}, and our results are robust to these small shifts, so we have not applied any changes (other than to scale to a common IMF).

We will consider five different \BH\ samples\footnote{We do not show results from \citet{KormendyHo} because their photometry is not as accurate as the others and they do not provide effective radii.} : those of \citet{Savo15}, \citet{Laesker14}, \citet{McConn13}, \citet{Beifiori12}, and \citet{Saglia16}.  The other five samples are based on the same sample of local galaxies with dynamical mass measurements of the central \BH, but with different estimates of the host galaxy velocity dispersion and luminosity.

The sample of \citet{Savo15} is the largest, most up-to-date set of galaxies with dynamically measured \BHs.  It consists of 66 galaxies with dynamical estimates of their black hole masses as reported by \citet{Graham13} or \citet{Rusli13}. Using $3.6\mu$ (Spitzer satellite) images, \citet{SavorgnanGrahamFits} modelled the one-dimensional surface brightness profile (measured along the major-axis and also the equivalent-circularized axis, i.e. the ``circularized'' profile) of each one of these 66 galaxies and estimated the structural parameters of their spheroidal component by simultaneously fitting a S\'{e}rsic function (used to describe the spheroid) in combination with additional components such as bars, discs, rings, nuclei. When available, kinematic information was used to confirm the presence and radial extent of rotating discs in the early-type galaxies.

Galaxy luminosities were converted into stellar masses assuming a Chabrier IMF and adopting a constant mass-to-light ratio of $(M/M_\odot)/(L/L_\odot)=0.6$ from \citet[e.g.][]{Meidt14}. \citet{Savo15} also explored more sophisticated ways of computing stellar masses based on colours, finding similar results.
The total galaxy effective radii (measured along the major- and the equivalent-axis) were estimated from the one-dimensional cumulative distribution of light as a function of galaxy radius, i.e., by imposing that the observed surface brightness profile integrated from $R=0$ to $R=R_e$ equal half of the total brightness. To these galaxy radii we assign a typical average uncertainty of 0.1 dex.  Central velocity dispersions are all derived from Hyperleda. In the following we exclude from their original sample NGC3842, NGC4889, UGC3789, and IC2560 which do not have Hyperleda velocity dispersions. We also remove another 10 galaxies that \citet{KormendyHo} classify either as ongoing mergers or as having uncertain \BH\ mass estimates (see their Tables~2 and 3). Finally, we do not consider the four galaxies for which \citet{Savo15} report only upper limits to the total magnitude (see their Table 1). This limits the final sample to 48 galaxies, of which 37 are E--S0 galaxies.

The photometry characterizing the \citet{Laesker14} sample of 35 galaxies, selected among those available in the literature with ``secure'' dynamical black hole mass measurements, was determined from deep, high spatial resolution images obtained from the wide-field WIRCam imager at the Canada--France--Hawaii--Telescope, accompanied by dedicated sky subtraction and improved fitting techniques similar to those by \citet{SavorgnanGrahamFits}.  To make a closer comparison with the other black hole samples and SDSS galaxies we adopt as a reference their standard S\'{e}rsic plus exponential luminosities, but note that using their ``improved'' luminosities -- based on more complex fitting models that may include additional components other than bulge and disc -- does not alter our conclusions. We convert their $K$-band luminosities into stellar masses assuming a Chabrier IMF adopting an average standard mass-to-light ratio of $(M/M_\odot)/(L/L_\odot)=0.67$ \citep[e.g.,][]{Longhetti09}. Velocity dispersions are all taken from Hyperleda.

From the original sample of \citet{McConn13} we retain only those objects which \citet{KormendyHo} label as secure, and further restrict to those with $3.6\mu$ luminosities and effective radii derived from S\'{e}rsic plus exponential fits by \citet{Sani11}. This reduces the original sample to 34 galaxies, of which 26 are E--S0s. We adopt their velocity dispersions obtained from integration of the spatially resolved measurements of the line-of-sight velocity dispersion and radial velocity from the sphere of influence of the black hole to one effective radius. The latter definition can reduce the values of central velocity dispersion by 10\%-15\% but, according to \citet{McConn13}, more accurately reflects the global structure of the host galaxy and is less sensitive to angular resolution.

The structural parameters in \citet{Beifiori12} are also homogeneously derived from bulge-to-disc decompositions of SDSS $i$-band images. Stellar masses were derived from adopting the mass-to-light ratio versus colour relations by \citet{Bell03SEDs}, who in turn adopted a ``diet'' Salpeter IMF, which yields about 0.15 dex higher stellar masses than a Chabrier IMF \citep[e.g.,][]{Bernardi10}. Stellar velocity dispersions come either from \citet{Beifiori09} or \citet{Gultekin09} and are rescaled to a velocity dispersion \sis\ equivalent to an effective stellar dispersion, measured within a circular aperture of radius $R_e$. When showing correlations with black hole mass, we will restrict to the subsample of galaxies by \citet{Beifiori12} with updated black hole masses from \citet{KormendyHo}.

To couple these datasets with SDSS galaxies, we convert SDSS velocity dispersions from $R_e/8$ to $R_e$ using the mean aperture corrections in \citet{Cappellari06}:
\begin{equation}
\left(\frac{\sigma_R}{\sigma_e} \right) = \left(R/R_e\right)^{-0.066}\, .
\label{eq|Cappellari}
\end{equation}
When dealing with the velocity dispersions $\sigma_{\rm HL}$ from the Hyperleda database \citep{Paturel03}, in which all measurements have been homogenized to a common aperture of 0.595 kpc, we also correct according to \eq\ref{eq|Cappellari}. These corrections are relatively small, and are not crucial for our results. The aperture correction in \eq\ref{eq|Cappellari} is consistent with other independent works \citep[e.g.,][]{Jorgensen96}. \citet{Cappellari13} claim a slight mass-dependent aperture correction, as expected in pressure-supported systems \citep[][]{Graham97}, though still, on average, in good agreement with \eq\ref{eq|Cappellari}.

While our work was being reviewed for publication, \citet{Saglia16} reported results from the SINFONI black hole survey.  For completeness, we briefly report results derived from their sample in \sect\ref{subsec|ImpactSpirals} and Appendix~A.
Bulge luminosities and half-light radii provided with this sample are determined from photometric decompositions that include bulges, discs, bars and rings. Bulge luminosities are then converted to stellar masses via dynamically determined mass-to-light ratios, and velocity dispersions are computed as line-of-sight weighted means within one effective radius. We remove from this sample 11 galaxies classified as unreliable by \citet{KormendyHo}.

\begin{figure*}
    \center{\includegraphics[width=15truecm]{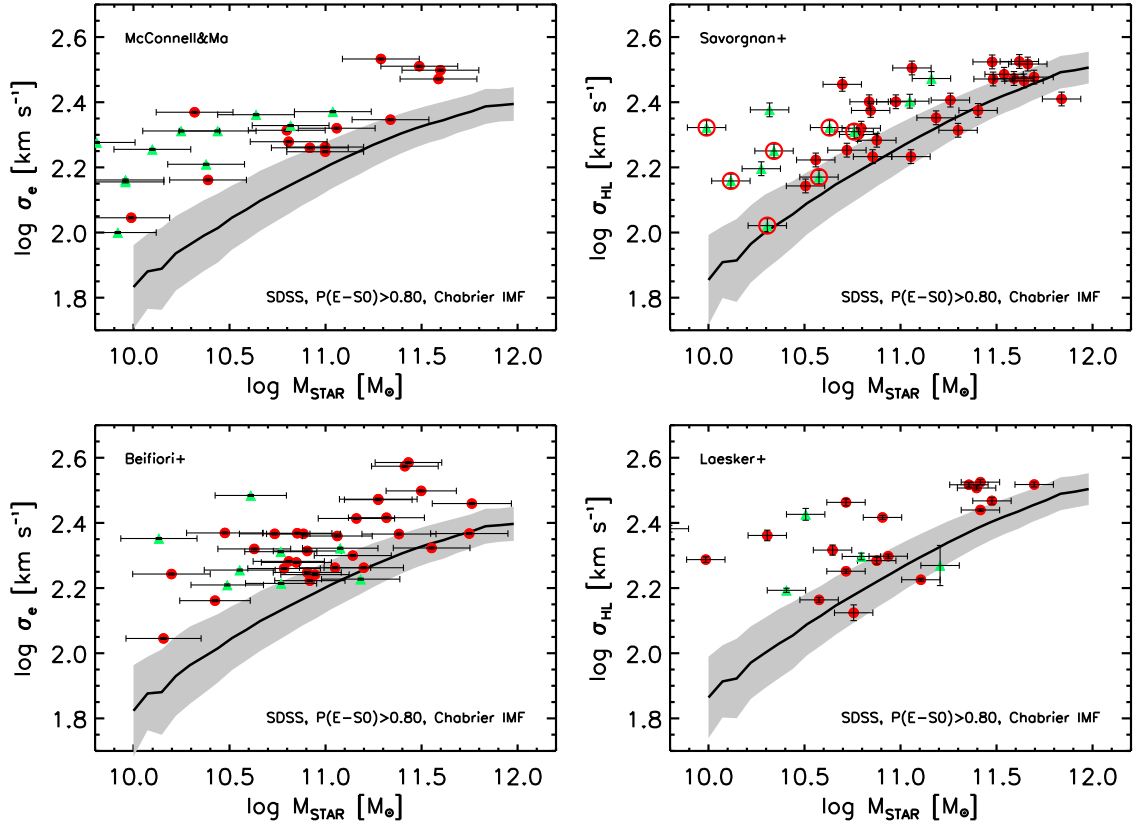}
    \caption{Mean velocity dispersion \sis\ at a given aperture (solid lines with gray bands), as labelled on the $y$-axis, as a function of the total stellar mass of SDSS galaxies with a probability $p$(E--S0) $>0.80$ of being classified as ellipticals and/or lenticulars (see text for details). The solid line in each panel shows the mean relation in the SDSS, based on the \SerExp\ stellar masses of \citealt{Meert15}; gray band shows the dispersion around the mean. The symbols show the local E--S0 galaxies with dynamically measured \BH\ masses from \citet[][top, left]{McConn13}, \citet[][top right]{Savo15}, \citet[][bottom left]{Beifiori12}, and \citet[][bottom right]{Laesker14}.  Filled red circles in each panel show ellipticals; green triangles show lenticulars. Open circles in the upper right panel mark the galaxies classified as barred by \citet{Savo15}. In all panels, most black hole hosts lie above the relations defined by the local population of SDSS galaxies regardless of morphological type.
    \label{fig|SigmaMstar}}}
\end{figure*}

\section{Selection bias}
\label{sec|selection}

\cite{Bernardi07}, and more recently \citet{Remco15}, noted that the scaling relations defined by the early-type galaxy hosts for which dynamically measured \BH\ masses are available differ from those of the early-type population as a whole:  \BH\ hosts tend to have larger velocity dispersions than early type galaxies of the same luminosity. \figu\ref{fig|SigmaMstar} shows that this bias is still present in the four more recent compilations/determinations (different panels) described in the previous section.  The solid line with grey bands shows the velocity dispersion \sis-total stellar mass \mstar\ relation of SDSS galaxies having probability $p$(E--S0) $>0.80$ of being classified as ellipticals and/or lenticulars according to the Bayesian automated classification of \citet{Huertas11}.  The symbols in each panel show the E--S0 galaxies with dynamically measured \BH\ masses:  in all panels, they lie significantly above the relation defined by the full SDSS.  We note that the SerExp decompositions assign larger luminosities to the galaxies with the highest velocity dispersions \citep[e.g.][]{Bernardi13, Bernardi12, Bernardi16}, thus further curving the \sis-\mstar\ relation (and related bias) with respect to previous estimates based on deVaucouleur's luminosities \citep[e.g.,][]{Tundo07,Bernardi10,Bernardi11a}.

\citet{Graham08FP} argued that the bias discussed by \cite{Bernardi07} was almost entirely due to lenticular and/or barred galaxies.  However, an error in Figure~7 of that paper invalidates this conclusion.  To double-check, the orange symbols in each panel of our \figu\ref{fig|SigmaMstar} show lenticulars:  the larger bias is evident in the \citet[][upper left]{McConn13} and \citet[][upper right]{Savo15} samples, but is less obvious in the bottom right panel.  Even if these objects are excluded, there is a clear offset from the relation defined by the SDSS galaxies. Indeed, in the \citet{Beifiori12} sample (bottom left) we have excluded all barred galaxies, and still find a clear offset. The offsets are evident whatever the exact sample considered, the selection adopted, the possible differences in estimating stellar masses in each subsample, and the aperture within which the velocity dispersion was estimated.  In \sect\ref{subsec|ImpactSpirals} we show there is also a clear offset if one considers bulge instead of total stellar masses (the \SerExp\ decompositions provide B/T estimates for the SDSS sample).

If the offset is a physical effect -- only the densest galaxies host black holes \citep[e.g.,][]{Saglia16}-- then it compromises the fundamental assumption in black hole demographic studies based on proxies:  that all galaxies host black holes.  However, there is a well-known selection effect:  \BH\ dynamical mass estimates are only possible if (some multiple of) the black hole's sphere of influence\footnote{Strictly speaking, the black hole sphere of influence depends on the exact stellar profile of the host galaxy and should thus be defined as the radius where $G\mbhe/r = G\mstare(<r)/r$. For a singular isothermal sphere, $G\mstare(<r)/r=\sigma^2$ for all $r$, so this yields \eq\ref{eq|rinfl}.  We have checked, however, that increasing or decreasing $r_{\rm infl}$ by a factor of $3$ has a relatively minor impact on our conclusions. Significantly larger values of $r_{\rm infl}$ would tend to reduce the bias we study in this paper -- but they are not supported by observations.},
\begin{equation}
 r_{\rm infl}\equiv G\mbhe/\sigma^2,
 \label{eq|rinfl}
\end{equation}
has been resolved \citep[e.g.,][]{Peebles72,MerrittFerrareseProceedings,Barth04,Bat10,Gultekin11,Graham13}. Only within $r_{\rm infl}$ (Keplerian) dynamics is expected to be dominated by the black hole \citep[e.g.,][]{MerrittFerrarese01}, and not adequately resolving the sphere of influence could significantly bias black hole mass estimates \citep[e.g.,][]{Merritt13}. The next section explores the consequences of this selection effect.

\section{Probing black hole-galaxy correlations and residuals through targeted Monte Carlo tests}
\label{sec|comparedataandsimulations}

We now describe the results of Monte Carlo simulations we have performed to study how the requirement that
\begin{equation}
 \theta_{\rm infl} \equiv r_{\rm infl}/d_{\rm Ang}
 \label{eq|thetainf}
\end{equation}
where $d_{\rm Ang}$ is the angular diameter distance, must exceed some critical angle $\theta_{\rm crit}$, impacts \BH\ and \BH-host scaling relations.  To illustrate our results, we set $\theta_{\rm crit}=0.1''$, a characteristic resolution limit for space-based (Hubble space telescope) observations.  We have verified that none of our conclusions is significantly changed if we increase the critical radius to, say, $\theta_{\rm crit}=0.5''$, which is more typical for ground-based measurements, Of course, increasing $\theta_{\rm crit}$ decreases the number of detectable objects.  In addition, the bias does not scale linearly with $\theta_{\rm crit}$ so a weak trend with resolution is expected.  Finally, we stress that this may not be the only selection effect in real samples; our goal is to study this effect in isolation.

\subsection{Setting up the simulations}
\label{subsec|simulations}

\begin{figure*}
    \center{\includegraphics[width=17.5truecm]{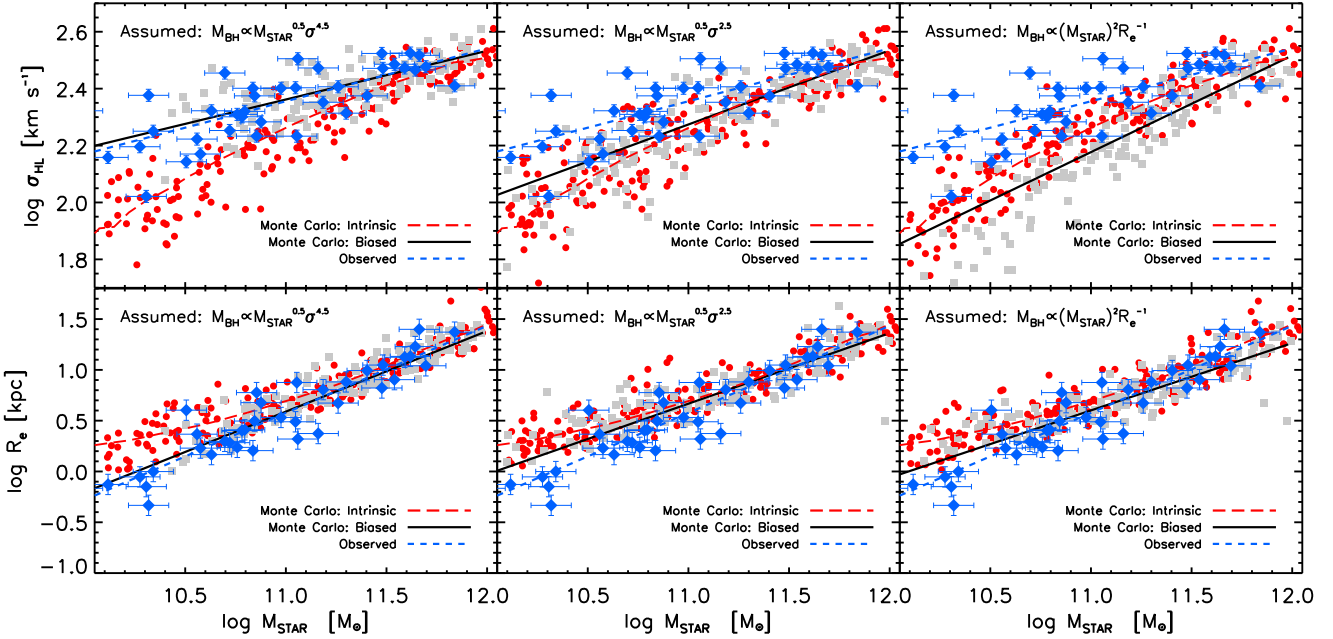}
    \caption{Host galaxy velocity dispersion (top) and effective radius (bottom) as a function of total stellar mass in Models~I (left), II~(middle) and III~(right) for which $\mbhe\propto\mstare^{0.5}\sigma^{4.5}$, $\mstare^{0.5}\sigma^{2.5}$ and $\mstare^2/R_e$, respectively.  Red circles and grey squares show a random subsample of 200 objects from the full sample, and the subset which is biased by the requirement that $r_{\rm infl}\ge 0.1''$, respectively.  Long-dashed red lines show the intrinsic relations in the full sample; solid black lines show linear fits to the selection biased subsample.  Blue diamonds with error bars show the dataset of \citet{Savo15}, and dashed blue lines show the associated straight line fits.}
\label{fig|BiasedFJrels}}
\end{figure*}

Our simulations, which follow the approach of \citet{Bernardi07}, work as follows:
\begin{enumerate}
  \item A comoving distance $d_{\rm Com}$ is drawn from a distribution which is uniform in comoving volume out to $200$~Mpc\footnote{This value was chosen to broadly  match the outermost distance for dynamical measurements of black holes (e.g., Cygnus A, \citealt{KormendyHo}). Reducing it to 100-150~Mpc does not qualitatively change any of our conclusions.}.  This cutoff is small enough that the difference between $d_{\rm Ang}$ and $d_{\rm Com}$ is irrelevant.
  \item A (total) stellar mass $\mstare$ is assigned from the \citet{Bernardi13} stellar mass function of ellipticals+lenticulars.
 \item A velocity dispersion is determined by drawing from a Gaussian distribution with mean and scatter as derived from the $\sigma-\mstare$ relation in the SDSS shown in the right panels of \figu\ref{fig|SigmaMstar}.
  \item The galaxy effective radii are set equal to those of the SDSS galaxy with the closest $\mstare$ and \sis.
\item Finally, a black hole mass is assigned to each galaxy in one of the following three ways (we discuss other possibilities in \sect\ref{subsec|MultipleRuns}).  In Models~I and~II,
      \begin{equation}
      \log \frac{\mbhe}{\msune}=
       \gamma + \beta \log\left(\frac{\sigma}{200\, {\rm km\, s^{-1}}}\right) + \alpha\log\left(\frac{\mstare}{10^{11}\, \msune}\right)\, ,
        \label{eq|MbhSigma}
      \end{equation}
with $(\gamma,\beta,\alpha)=(7.7,4.5,0.5)$ for \modI\ and $(\gamma,\beta,\alpha)=(7.75,2.5,0.5)$ for \modII.
      In \modIII,
      \begin{equation}
       \log \frac{\mbhe}{\msune} = \gamma + \beta\log\left(\frac{R_e}{5~{\rm kpc}}\right) + \alpha\log\left(\frac{\mstare}{10^{11}\, \msune}\right) \, ,
       \label{eq|MbhW}
      \end{equation}
with $(\gamma,\beta,\alpha)=(7.4,-1,2)$.  For all three models, we add 0.25~dex rms (Gaussian) scatter around the assumed mean relation.
  \item We repeat the steps above many times to create what we call the full \BH\ sample.
  \item For each object in the full sample we define $\theta_{\rm infl}$ following equation~(\ref{eq|thetainf}).  The subset of objects with $\theta_{\rm infl} \ge \theta_{\rm crit}$ make up our selection-biased sample.
\end{enumerate}

\modII\ was chosen because it is similar to the observed (selection biased) scaling reported in the literature \citep[e.g.,][]{Hop07FP,KormendyHo,McConn13}.  \modI\ has a stronger intrinsic dependence on \sis\ which we argue later is required to explain all the observed correlations.  And \modIII\ was chosen mainly because it scales like the potential energy, so $\sigma$ does not play a fundamental role; rather, in this model, the \mbh-\sis\ correlation is a result of more fundamental correlations with \mstar\ and $R_e$.  While our choices for the intrinsic scatter are close to those reported in the literature, for reasons that will become clear later, the normalizations $\gamma$ in all the three models above are $\sim 0.4-0.6$~dex ($\sim 2.5-4\times$) lower than the values given in the literature.

Notice that we do not distinguish between intrinsic scatter and observational errors in our Monte Carlo simulations:  we return to this later.  In addition, while all the mock-based results that follow are presented in terms of total stellar mass, using bulge stellar masses (and radii) instead yields qualitatively similar results (see \sect\ref{subsec|ImpactSpirals}).

\subsection{The selection biased \sis-\mstar\ and $R_e$-\mstar\ relations}
\label{subsec|RelationsScatters}

\begin{figure*}
    \center{\includegraphics[width=16truecm]{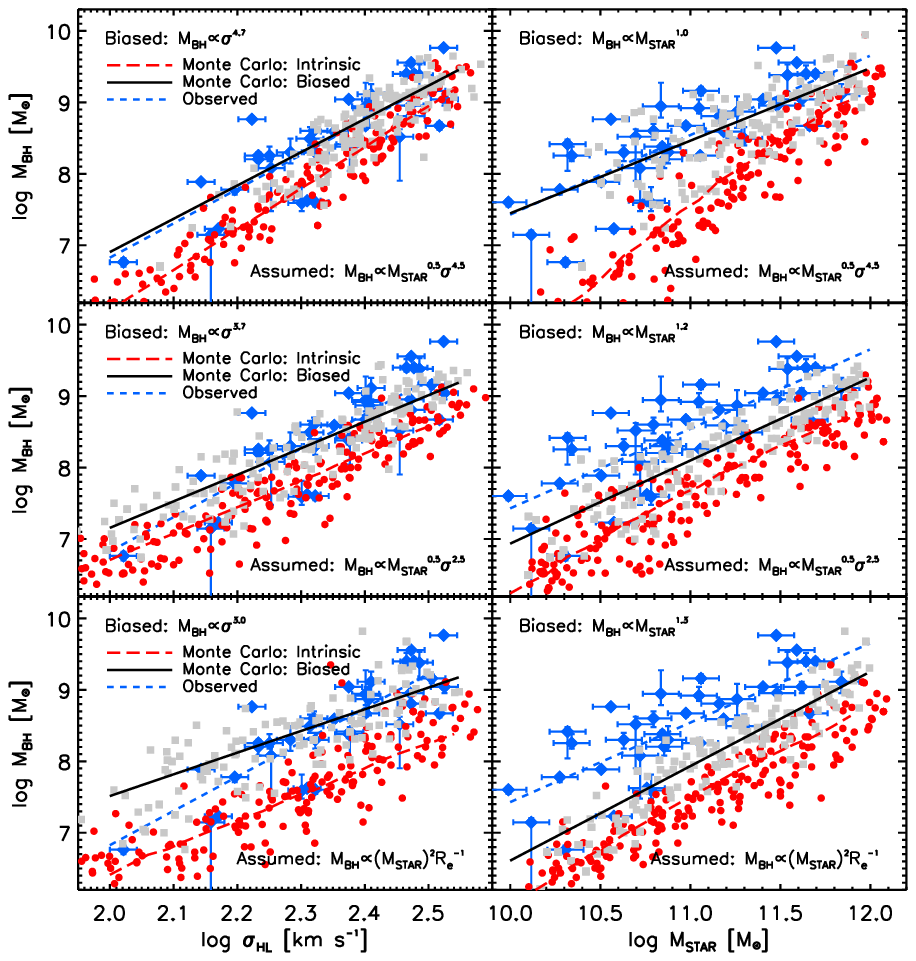}
    \caption{Black hole scaling relations:  \mbh-\sis\ (left) and \mbh-\mstar\ (right), in \modI\ ($\mbhe \propto \mstare^{0.5}\sigma^{4.5}$, top), \modII\ ($\mbhe \propto \mstare^{0.5}\sigma^{2.5}$, middle), and \modIII\ ($\mbhe \propto \mstare^2/R_e$, bottom) in the same format as the previous figure.  Red circles and grey squares show a random subsample of 200 objects from the full and selection biased subsamples, respectively.  Red long-dashed and black solid curves show the associated mean values of \mbh\ as \sis\ and \mstar\ vary.  Resolving the black hole sphere of influence biases the observed relations so that they lie significantly above the intrinsic ones; they overestimate \mbh\ by factors of $3\times$ or more.  Blue symbols with error bars show the \citet{Savo15} dataset which is only really matched by the (selection biased) \modI.\label{fig|ScalingsMocks}}}
\end{figure*}

In this section we answer the basic question: can the $r_{\rm infl}$-selection effect help explain the discrepancy shown in \figu\ref{fig|SigmaMstar}?  We will use ``scaling relations'' to address this.  In all cases, this means we treat the quantity plotted on the $y$-axis as the dependent variable when fitting.  We never treat it as the independent variable, nor do we perform `bisector'-like fits.  When fitting straight lines to the data, we have compared three different linear regression algorithms finding very similar results.  The values we report have been performed with the IDL routine \texttt{robust\_linfit}.

In the next two figures, red circles and grey squares show 200 randomly chosen members of the intrinsic and selection biased Monte-Carlo samples, and blue symbols with error bars show the E+S0s in the \citet{Savo15} dataset.  Long-dashed red curves show the intrinsic scaling relations, solid black lines show linear fits to the selection biased sample, and short-dashed blue lines show linear fits to the data.

The top panels of \figu\ref{fig|BiasedFJrels} show the \sis-\mstar\ relation:
left, middle, and right-hand panels show results for Models~I, II and~III.
There is a clear offset between the intrinsic and selection biased objects in the top left panel, a smaller one in the top middle, and a bias in the opposite sense in the top right panel.  This is easy to understand:  In \modII, $r_{\rm infl}\propto (\mstare\sise)^{0.5}$ is nearly a function of \mstar\ only (the range of \mstar\ values is much broader than of \sis).  Correlations with the variable on which the selection was made will be unbiased, and, since the correlation shown is at fixed \mstar, the $r_{\rm infl}$ selection does not bias the \sis-\mstar\ relation in the middle panel very much.  However, in \modI, $r_{\rm infl}\propto \mstare^{0.5}\sise^{2.5}$ is nearly a function of \sis\ only, so requiring $\theta_{\rm infl}\ge\theta_{\rm crit}$ will tend to select large \sis, which is what we see in the left hand panel.  In contrast, $r_{\rm infl}\propto (\mstare^2/R_e)/\sigma^2$ in \modIII, so $\theta_{\rm infl}\ge\theta_{\rm crit}$ tends to select small \sis\ in the right hand panel.

Comparison with the blue symbols in the top panels shows that \modI\ is remarkably similar to the data, whereas Models~II and~III are not.  The discrepancy between the \sis-\mstar\ relation in the selection-biased sample and the data (i.e., the \citealt{Savo15} E+S0s) is most pronounced in \modIII, because it has no \sis\ factor in \mbh\ to cancel the $\sigma^2$ factor in the definition of $r_{\rm infl}$, so the selection biased sample is composed of objects with smaller \sis\ (rather than larger) for their \mstar.

\begin{figure*}
    \center{\includegraphics[width=15truecm]{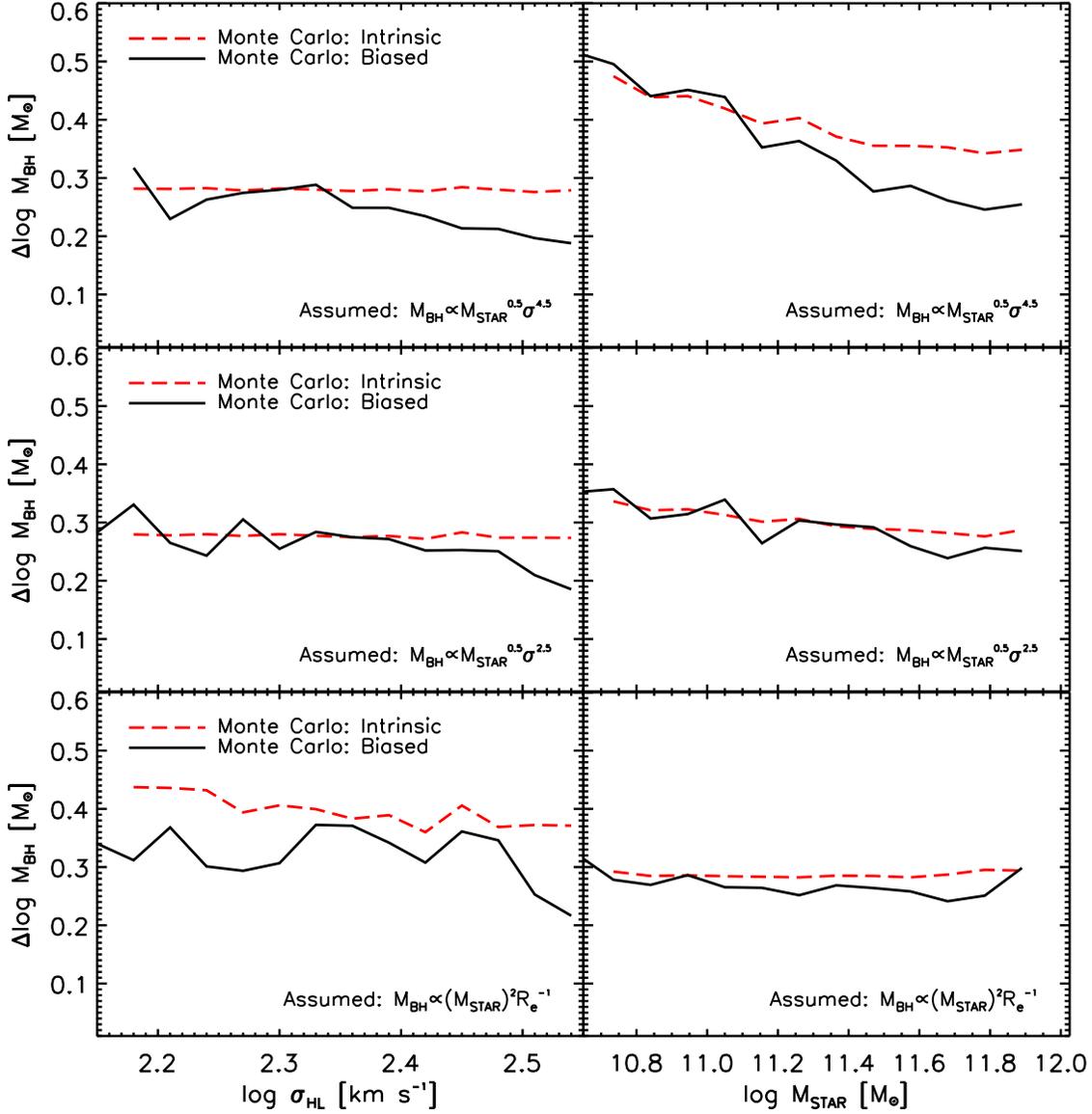}
    \caption{Scatter around the mean relations for the mock catalogues shown in \figu\ref{fig|ScalingsMocks}. Long-dashed red lines mark the intrinsic scatter around the mean relations; solid black lines show the scatter in the selection-biased subsamples. The decreasing scatter in the \mbh-\mstar\ relation for \modI\ (upper right), which is amplified in the selection biased sample, is a direct consequence of the fact that the scatter around the mean \sis-\mstar relation decreases at large \mstar\ (\figu\ref{fig|SigmaMstar}). \label{fig|ScattersMocks}}}
\end{figure*}

For completeness, the bottom panels of \figu\ref{fig|BiasedFJrels} show a similar analysis of the $R_e$-\mstar\ relation.  \modII\ is nearly unbiased by the selection effect for the same reason as before; and while there is a small bias (to slightly smaller $R_e$) in models I and~III, it is much smaller than for \sis-\mstar.  \footnote{The slight bias can be understood in terms of the virial theorem:  at fixed \mstar, large \sis\ means smaller $R_e$, and we know that the selection effect in \modI\ favours large \sis.  The bias appears small because the intrinsic $R_e$-\mstar\ relation is tighter, i.e. has less scatter, than the \sis-\mstar\ relation.}  All the models are in reasonable agreement with the \citet{Savo15} data.

\begin{figure*}
    \center{\includegraphics[width=15truecm]{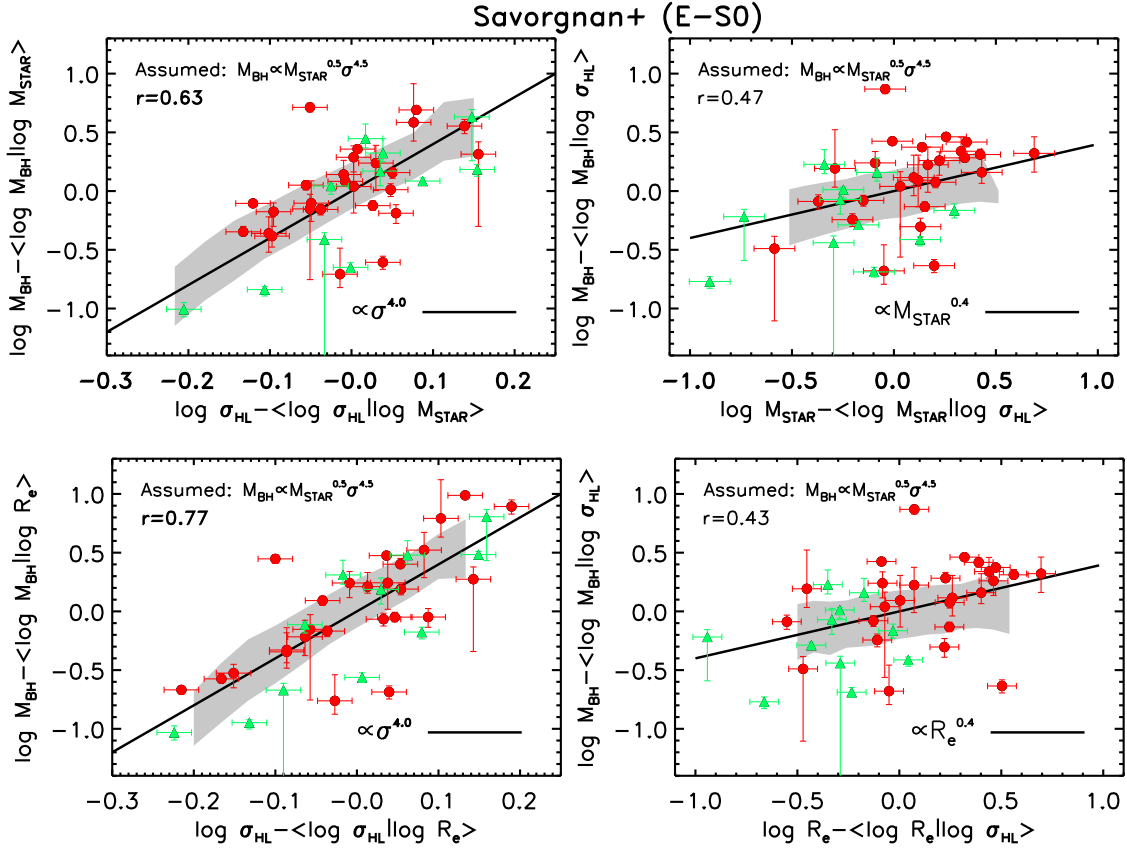}
    \caption{Correlations between residuals from the observed scaling relations, as indicated.  Red circles and green triangles show E and S0 galaxies from the \citet{Savo15} E--S0 sample, while grey bands show the corresponding measurements in our (selection biased) \modI.  The selection bias tends to reduce the slope of the \mbh-\sis\ relation (from $\beta=4.5$ to $\sim 4$), so solid lines in panels on the left show a slope of 4.  In the panels on the right, solid lines with a slope of 0.4 approximate the expected scalings with stellar mass or effective radius in the biased samples. Correlations with velocity dispersion appear to be stronger than those with the other two quantities, in good agreement with \modI.
    \label{fig|MockResidualsModI}}}
\end{figure*}

\subsection{Selection-biased \mbh\ scaling relations}
\label{subsec|RelationsScatters}

Having shown that the selection biased samples are similar to the data -- with \modI\ faring better than the other two models for \sis-\mstar\ (in general, the dependence on velocity dispersion, labelled by the slope $\beta$, must be large to explain the observed offset in the \sis-\mstar\ relation) -- and that the selection biased \modI\ sample is biased compared to the intrinsic \sis-\mstar\ relation --  we now turn to correlations with \mbh.

\figu\ref{fig|ScalingsMocks} shows the \mbh-\sis\ (left) and \mbh-\mstar\ (right) relations
in our Monte Carlo simulations based on \modI\ (top), \modII\ (middle), and \modIII\ (bottom).
All models predict biased scaling relations that have higher normalizations and in some cases flatter slopes than the intrinsic ones.  This is the main reason why we chose lower normalization factors $\gamma$ for all our Monte-Carlo models (cfr. \sect\ref{subsec|simulations}).  This bias becomes more pronounced as the input slope $\beta$ or the input scatter increase (also see discussion of \figu\ref{fig|MultipleRuns}).  This is why the bias induces a stronger upwards boost in the \mbh-\mstar\ relation for \modI\ than \modII\ or~III (right panels).

In addition, notice the curvature in the intrinsic \mbh-\mstar\ relation (long-dashed red lines in the panels on the right), which is most evident for \modI. Since \eq\ref{eq|MbhSigma}, which we used to generate \mbh, assumes pure power-law relations, this curvature is entirely a consequence of curvature in the \sis-\mstar\ relation (\figu\ref{fig|SigmaMstar}).  Curvature in the observed (selection biased) \mbh-\mbulge\ relation has been reported by \citet{Graham13}.  Our analysis suggests that this curvature is due to galaxy formation physics, and need not imply anything fundamental about black hole formation or mergers \citep[see also][]{FF15}. The intrinsic \mbh-\mstar\ relation in \modI\ can be approximated by
\begin{multline}
\log \frac{\mbhe}{\msune} =
 7.574 + 1.946\,\log \left(\frac{\mstare}{10^{11}\msune}\right)- 0.306\\
\times\left[\log\left(\frac{\mstare}{10^{11}\msune}\right)\right]^2
 - 0.011\,\left[\log\left(\frac{\mstare}{10^{11}\msune}\right)\right]^3 ,
 \label{eq|IntrinsicMbhMstar}
\end{multline}
while the intrinsic \mbh-\sis\ relation is
\begin{equation}
\log \frac{\mbhe}{\msune}=7.8+5.7\log\left(\frac{\sigma_{HL}}{200\, {\rm km\, s^{-1}}}\right).
\label{eq|IntrinsicMbhSigma}
\end{equation}
The first reflects the curvature in the \sis-\mstar\ relation, and the second is consistent with the \mstar-\sis\ relation having a linear scaling of the type $\mstare\propto\sise^{2.4}$, at least below $\sise \lesssim 260\, {\rm km\, s^{-1}}$. The Milky Way to date is the best resolved dynamical measurement of a central black hole with $\mbhe\sim 4.3\times10^6\, \msune$, $\sise \sim 200\, {\rm km\, s^{-1}}$ \citep[][]{Scott13} and stellar mass of $\log \mstare \sim 10.7$ \citep[e.g.,][for a Chabrier IMF]{Lic15}. It is also awkwardly renown to be a strong outlier with respect to the observed scaling relations especially with stellar mass \citep[e.g.,][]{Marconi03}, but it is instead fully consistent with our intrinsic relations of \modI.

Having illustrated the bias in the mean relations, we now consider the scatter around the relations, computed in our mocks as the 1\sis\ dispersion around the mean.  The six panels in \figu\ref{fig|ScattersMocks} are for the same relations shown in \figu\ref{fig|ScalingsMocks}:  the long-dashed, red lines show the scatter in the full sample (i.e., the scatter around the long-dashed red lines in \figu\ref{fig|ScalingsMocks}), and the black solid lines show the scatter measured in the selection biased subsample (the scatter around the black lines in \figu\ref{fig|ScalingsMocks}).  In all models, the scatter in the biased samples is comparable to, or as much as $\sim 30\%$ smaller than, the intrinsic scatter. Although the data are too sparse to allow a reliable determination of the scatter, they do show a tendency to decrease at large masses which is in qualitative agreement with our simulations.

It is sometimes argued that because the observed scatter around the \mbh-\mstar\ relation is of the same magnitude as that around the \mbh-\sis\ relation, especially in early-type, massive galaxies, it is plausible that the \mbh-\mstar\ relation is at least as, if not more, fundamental.  However, the top panels show that this argument is flawed because in \modI\ velocity dispersion is more important than stellar mass:  the scatter around the observed \mbh-\mstar\ relation seems comparable to the one on the left panel, especially at large masses, because of the selection effect.  In addition, some groups have reported a tendency for the scatter to decrease at large masses in models characterized by repeated \BH-\BH\ mergers \citep[e.g.,][]{Peng07,JahnkeMaccio,Hirschmann10}.  \figu\ref{fig|ScattersMocks} suggests that such arguments should be treated with caution, as this trend, clearly observed in \modI\ (upper panels), entirely reflects the decrease in scatter in the velocity dispersion with increasing stellar mass (\figu\ref{fig|SigmaMstar}) which the selection bias amplifies (solid line). \citet{Graham13} have addressed this point differently, by arguing that the low-mass end of the \mbh-\mbulge\ diagram does not converge to a relation with a slope of unity, as required in the many-merger scenario.

\begin{figure*}
    \center{\includegraphics[width=15truecm]{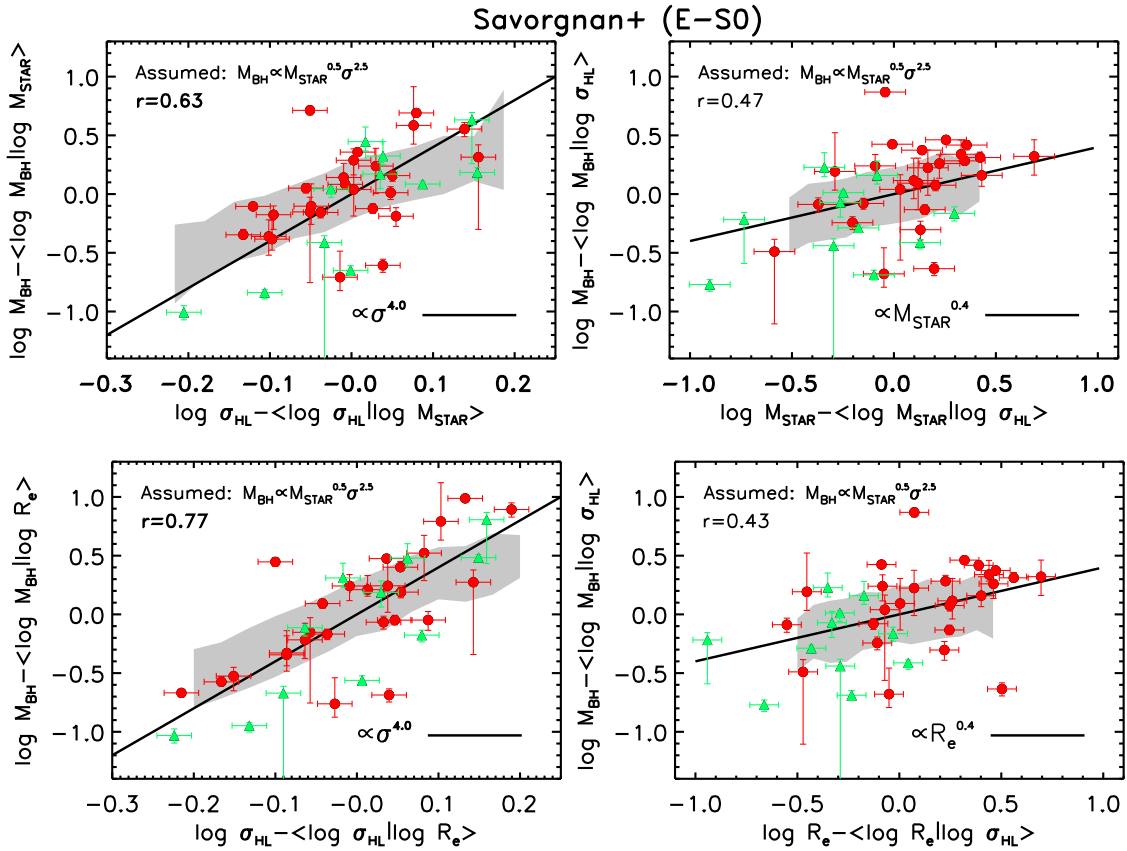}
    \caption{Same as \figu\ref{fig|MockResidualsModI}, but with grey bands now showing our (selection biased) \modII.  The dependence on \sis\ (left hand panels) is weaker than in the data.
     \label{fig|MockResidualsModII}}}
\end{figure*}

\subsection{Selection-biased black hole demographics}
\label{subsec|selectiondemo}

The selection effect has another important consequence.  If one uses the observed \sis-\mstar\ (or bulge mass) relation to translate between \sis\ and \mstar, then the observed \mbh-\sis\ relation predicts a factor of $\sim 3$ lower black hole masses than the observed \mbh-\mstar\ relation \citep{Tundo07,Bernardi07,Graham07BHMF}.  Accounting for the scatter around these relations does not resolve the discrepancy \citep{Tundo07}.  The selection-biased relations (black lines in \figu\ref{fig|SigmaMstar}) in our Monte Carlo simulations result in a similar discrepancy.  For example, in \modI, $\log(\mstare/\msune)=11$ would predict a black hole mass of $\log(\mbhe/\msune) \sim 8.4$ (upper right).  \figu\ref{fig|SigmaMstar} (right panels) shows that $\log(\mstare/\msune)=11$ corresponds to $\log(\sigma_{\rm HL}/{\rm km~s}^{-1}) \sim 2.25$, for which the solid line in the upper left panel of \figu\ref{fig|ScalingsMocks} suggests $\log(\mbhe/\msune) \sim 7.9$.  This discrepancy is somewhat smaller, but still present, in the other two models.  Accounting for the scatter around these selection-biased relations does not resolve the discrepancy. These are selection effects: in all models the discrepancy is much smaller if one uses the intrinsic relations (long-dashed red curves), and it disappears entirely (by definition) if one accounts for the intrinsic scatter.

In this context, the results from the cosmological black hole model presented by \citet{Sija15} and developed in the framework of the Illustris simulation \citep{Voge14} are informative. When normalized to the (biased) black hole mass-bulge stellar mass relation, their model systematically overproduces the \mbh-\sis\ relation by a factor of $\gtrsim 3$ at $\log(\sigma/{\rm km~s}^{-1}) \gtrsim 2.3$ (their \figu6).  Our work suggests this is a consequence of normalizing to a relation that has been biased by selection effects, rather than to the intrinsic relation.

\subsection{Correlations between residuals}
\label{subsec|MockResiduals}

Studying correlations between the residuals from various scaling relations is an efficient way of determining if a variable is fundamental or not.  For example, if \mbh\ is determined by \sis\ alone (e.g., if $\alpha=0$ in \modI) then residuals from correlations with \sis\ should be uncorrelated.  In this case, residuals from the \mbh-, \mstar-, and $R_e-$\sis\ relations should not correlate with one another \citep[e.g.,][]{Bernardi05,ShethBernardi12}.  In contrast, not only should residuals from the \mbh-\mstar\ relation correlate with residuals from the \sis-\mstar\ relation, but the slope of this correlation between residuals should be the same as that of the \mbh-\sis\ relation itself; this is what indicates that \sis\ controls the \mbh-\mstar\ correlation.

The grey bands in each panel of \figu\ref{fig|MockResidualsModI} show residuals along the $y$-axis from the scaling relations measured in the selection biased subsamples of \modI.  In this Model, residuals from the true intrinsic correlations with \sis\ (left panels) should correlate with a slope of $\beta \sim 4.5$, and those as a function of stellar mass (or effective radius) with a slope of $\alpha \sim 0.5$.  This is indeed what the simulations show, though the selection bias tends to slightly flatten the slope of the \mbh-\sis\ relation (top left panel of \figu\ref{fig|ScalingsMocks}) from a slope of $\beta=4.5$ to $\beta \sim 4$ and also the biased $\alpha\sim 0.3-0.4$. This flattening would be even more pronounced for higher values of the intrinsic $\beta$ and/or scatter (see discussion of \figu\ref{fig|MultipleRuns}).

The red circles (ellipticals) and green triangles (S0s) in the same \figu\ref{fig|MockResidualsModI} show a similar analysis of the residuals from scaling relations in the observed E+S0 sample of \citet{Savo15}.  The slopes in all the panels on the left are $\beta\sim 4$.  We quantify the strength of each correlation using the Pearson correlation coefficient $r$, which we report in the top left corner of each panel.  These values show that the correlations on the right are weaker than those on the left, indicating that \sis\ is the more important of each pair.  This is important as it shows that the selection biased sample still correctly indicates that \sis\ is more fundamental.

In each panel, the grey band defined by \modI\ is consistent with the correlation measured in the \citet{Savo15} sample.  \figu\ref{fig|MockResidualsModII} shows a similar analysis of \modII; the grey bands in the left hand panels show that the dependence on \sis\ in this model is weaker than in the data.

Despite the small size of the dataset, we have attempted to quantify the uncertainties on the trends shown in the previous figures by using a bootstrapping technique.  We randomly remove three sources from the \citet{Savo15} sample, measure the correlations and hence the residuals from the correlations, record the slope and scatter for the correlations between residuals, and repeat 100 times.  For the Monte-Carlos, we instead generate 100 mock samples, each having 50 objects, for which we measure the slope and scatter of the correlations between residuals.

\begin{table*}
\caption{Mean slope and its uncertainty for the correlation between residuals named in the first column. The compact notation in the first column
has the meaning $\Delta(X|Y)=\log X-\langle \log X|\log Y \rangle$. The second, third, and fourth columns are for the $E-S0$s, all galaxies, and all bulges from \citet{Savo15}. The fifth, sixth, seventh, and eight columns are the corresponding results from the Monte Carlo simulations of \modI\ (total and bulge stellar masses), \modII, and \modIII.  \modI\ tends to be in better agreement with the data, though uncertainties are still substantial (see text for details).}
\label{Table|Slopes}
\begin{tabular}{|l|l|l|l|l|l|l|l|}
  \hline
  Residual                & \,\,\,\,\,\, E-S0 & \,\,\,\,\,\,\, All & \, All bulges & \, \modI\ & \, \modI\ (bulges) & \modII\ & \, \modIII\ \\
  \hline
  \,\,\,\,\,\, (1)                    &   \,\,\,\,\,\,\,\,  (2)              &   \,\,\,\,\,\,\,\,  (3)              &    \,\,\,\,\,\,\,\,\, (4)        &      \,\,\,\,\,\,\,\, (5)
  &  \,\,\,\,\,\,\, (6)            & \,\,\,\,\,\,\,\,(7)      &  \,\,\,\,\,\,\,\, (8)    \\
  \hline
$\Delta(\mbhe|\mstare)$ vs $\Delta(\sise|\mstare)$    &   3.24$\pm$0.64   &    4.42$\pm$0.58   &   3.60$\pm$0.51    &   3.68$\pm$1.30   &    3.97$\pm$1.40  &  2.47$\pm$0.80   &   0.44$\pm$1.02   \\
$\Delta(\mbhe|R_e)$  vs $\Delta(\sise|R_e)   $         &   3.94$\pm$0.47   &    4.70$\pm$0.43   &   3.86$\pm$0.43    &   4.45$\pm$1.00   &    4.99$\pm$1.05  &   3.18$\pm$0.53   &   2.73$\pm$0.83   \\
$\Delta(\mbhe|\sise)$ vs $\Delta(\mstare|\sise)$      &   0.54$\pm$0.14   &    0.34$\pm$0.16   &   0.42$\pm$0.09    &   0.35$\pm$0.24   &    0.34$\pm$0.25  &   0.47$\pm$0.24   &   1.22$\pm$0.44   \\
$\Delta(\mbhe|\sise)$  vs $\Delta(R_e|\sise)   $       &   0.45$\pm$0.14   &    0.31$\pm$0.15   &   0.41$\pm$0.09    &   0.54$\pm$0.32   &    0.34$\pm$0.24  &   0.38$\pm$0.26   &   0.23$\pm$0.53   \\
  \hline
\end{tabular}
\end{table*}

\begin{table*}
\caption{Same as Table~\ref{Table|Slopes}, but now the second, third, fourth, fifth, and sixth columns are from the $E+S0$s of \citet{Savo15}, \citet{Saglia16}, \citet{McConn13}, \citet{Laesker14}, and \citet{Beifiori12}.  The other samples are in good agreement with the \citet{Savo15} sample and with \modI\ (seventh column), suggesting, if anything, an even weaker dependence on stellar mass.}
\label{Table|SlopesData}
\begin{tabular}{|l|l|l|l|l|l|l}
  \hline
 Residual                & Savorgnan+ & \,\,\,\ Saglia+ & McConnell\&Ma & \, L\"{a}sker+ & Beifiori+ & \modI\ \\
 \hline
   \,\,\,\,\,\, (1)                    &   \,\,\,\,\,\,\,\,  (2)              &   \,\,\,\,\,\,\,\,  (3)              &    \,\,\,\,\,\,\,\,\, (4)        &      \,\,\,\,\,\,\,\, (5)
  &  \,\,\,\,\,\,\,  (6)  &  \,\,\,\,\,\,\,  (7)       \\
  \hline
 $\Delta(\mbhe|\mstare)$ vs $\Delta(\sise|\mstare)$   &    3.24$\pm$0.64    &   3.47$\pm$0.65   &    3.60$\pm$0.94   &    4.39$\pm$0.95 &   4.36$\pm$0.81 & 3.68$\pm$1.30   \\
 $\Delta(\mbhe|R_e)$ vs $\Delta(\sise|R_e)   $       &    3.94$\pm$0.47    &   3.72$\pm$0.52   &    4.16$\pm$0.71   &    4.63$\pm$0.89  &   3.95$\pm$0.62 & 4.45$\pm$1.00 \\
 $\Delta(\mbhe|\sise)$ vs $\Delta(\mstare|\sise)$     &    0.54$\pm$0.14    &   0.23$\pm$0.13   &    0.39$\pm$0.23   &    0.33$\pm$0.19 &   0.09$\pm$0.94 & 0.35$\pm$0.24  \\
 $\Delta(\mbhe|\sise)$ vs $\Delta(R_e|\sise)   $     &    0.45$\pm$0.14    &   0.26$\pm$0.12   &    0.39$\pm$0.22   &    0.39$\pm$0.24  &   0.05$\pm$0.28 & 0.54$\pm$0.32 \\
  \hline
\end{tabular}
\end{table*}

\begin{figure*}
    \center{\includegraphics[width=15truecm]{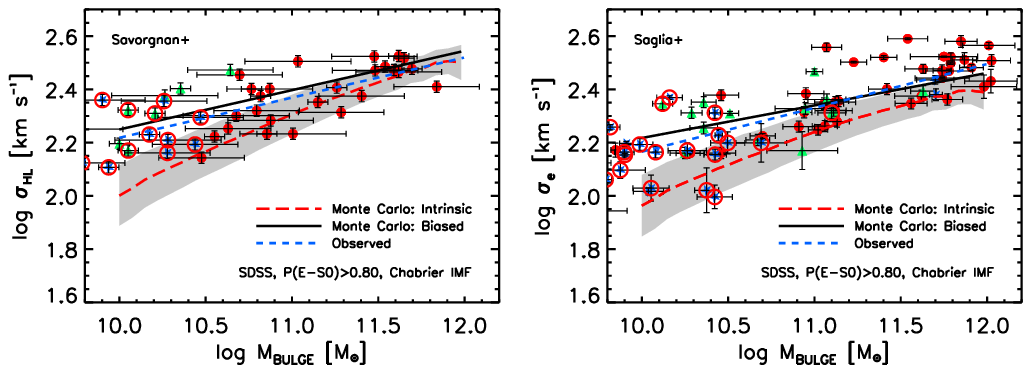}
    \caption{Left: Same as top right panel of \figu\ref{fig|SigmaMstar} but with \mbulge\ rather than total \mstar\ for the sample of \citet{Savo15}. Right: Same as left panel but for the sample of \citet{Saglia16}, for which we have used $\sigma_e$ as $\sigma_{\rm HL}$ is not available. We also include results of the \modI\ Monte Carlos when using only bulge stellar masses. Filled red circles, green triangles, and blue stars in both panels correspond, respectively, to ellipticals, S0, and the bulges of spirals. Open red circles mark the galaxies with bars. The predicted biased relations (black, solid lines) agree well with those observed (blue, dashed lines).
    \label{fig|SigmaMbulge}}}
\end{figure*}

The mean slopes and standard deviations over the 100 realizations are reported in Table~\ref{Table|Slopes}. The second, third, and fourth columns report the results of the bootstrapping on the data considering only E-S0, all galaxies, and all bulges, respectively, while the fifth/sixth, seventh, and eight columns are the results of \modI\ (total and bulge stellar masses), \modII, and \modIII, respectively. This shows that \modI-total stellar masses (column 5) provides slopes that are well consistent with those of the E-S0 \citet{Savo15} sample (column 2). However, the uncertainties on the slopes are relatively large, so even \modII, which tends to predict flatter slopes than what observed (\figu\ref{fig|MockResidualsModII}), is only discrepant at the $1-2\sigma$ level. \modIII, the residuals of which are shown in Appendix~B, appears to be more than $2 \sigma$ discrepant, especially in the residuals of velocity dispersion at fixed stellar mass (first row).

A similar analysis of the residuals in the \citet{Saglia16}, \citet{McConn13}, \citet{Laesker14}, and \citet{Beifiori12} samples is included in Appendix~A, with the results of the statistical analysis reported in Table~\ref{Table|SlopesData} (columns 3, 4, 5, and 6, respectively), compared to the \citet{Savo15} sample (column 2) and the predictions of \modI\ (column 7). For all samples in Table~\ref{Table|SlopesData} we restrict to E-S0 with total stellar masses and effective radii, except for the \citet{Saglia16} dataset, for which only bulge masses and radii are available (for this sample we also include non-barred spirals). Mean slopes and uncertainties are again computed from 100 bootstrap iterations in which 3 sources were removed at a time, except for the smaller \citet{Beifiori12} sample, for which we only removed a single object at a time.
These other samples show the same trends we found in the \citet{Savo15} sample.  If anything, the dependence on \sis\ is stronger, and that on \mstar\ or $R_e$ is weaker, so that \modI\ fares better than the others.

\subsection{The impact of spirals and bulge-to-total decompositions}
\label{subsec|ImpactSpirals}

In the previous section we noted that the residuals around the black hole-galaxy scaling relations suggest that velocity dispersion is the most important property of a galaxy with regards to the black hole at its centre. That analysis was based on a sample of early-type galaxies.  In this section we include the spirals from the \citet{Savo15} with ``secure'' black hole mass measurements as reported in \citet{KormendyHo}.  As for the early-types, velocity dispersions for these galaxies are taken from the Hyperleda data base and total half-light radii are derived as explained in \sect\ref{sec|data}.

In the context of including spirals, however, it is possible that the bulge mass is more relevant than the total.  We noted in \sect\ref{sec|data} that \mbulge\ is significantly more difficult to estimate reliably \citep[e.g.][]{Meert13}.  Nevertheless, \citet{Meert15} provide B/T decompositions for their S\'{e}rsic-Exponential reductions, which we have used to estimate \mbulge\ in the SDSS. \citet{Savo15} also considered detailed galaxy decompositions that take into account spheroid, discs, spiral arms, bars, rings, halo, extended or unresolved nuclear source and partially depleted core, and checked for consistency with galaxy kinematics \citep[e.g.,][]{Arnold14}. So, we have used bulge stellar masses to see how the offset in the top right panel of \figu\ref{fig|SigmaMstar} changes when we replace $\mstare\to\mbulgee$.  We continue to restrict the analysis to E-S0 SDSS galaxies, as determining the central velocity dispersion of spirals from the SDSS spectra (which are not spatially resolved) is not possible.
Adopting the E-S0 bulge sample as representative of the full galaxy population between $10^{10}<\mstare/\msune<10^{12}$ is a safe assumption given that the bulges of spirals have structural properties that follow the scaling relations of the bulges of early-type galaxies quite well \citep[e.g.,][]{Bernardi12}.

The left panel of \figu\ref{fig|SigmaMbulge} shows the \citet{Savo15} data with spirals included (blue stars), but using \mbulge\ rather than total masses.  It shows that there is a clear offset from the \sis-\mbulge\ relation of the SDSS, qualitatively like that seen in Figure~\ref{fig|SigmaMstar}.  The significantly larger error bars reflect the larger uncertainties in estimating \mbulge. Later-type galaxies show the largest offsets, in agreement with \citet{Graham08FP}.  Also note that the \citet{Meert15} reductions have a slight systematic tendency to set B/T~$\approx 0.9$ even when B/T~$=1$ \citep[see Figure~9 of][]{Meert13}.  Therefore, at large \mbulge\ where we expect to have B/T~$\to 1$, the SDSS relation is shifted slightly more to the left than it should be.  Removing this systematic would slightly increase the offset between the SDSS and the \citet{Savo15} sample.

The right panel of \figu\ref{fig|SigmaMbulge} compares the SDSS \sis-\mbulge\ relation with the bulge masses and velocity dispersions in the sample of \citet{Saglia16}. The offset between the two data sets here is slightly more pronounced than it was for the \citet{Savo15} data (on the left).  Note that the mass-to-light $M/L$ ratios adopted by \citet{Saglia16} are dynamical ones: they are not derived from spectral analysis. \citet{Saglia16} argue that their $M/L$ are broadly consistent with those obtained assuming a Kroupa \citep{Kroupa01} IMF which, if anything, should yield systematically larger stellar masses at fixed velocity dispersion than those obtained from a Chabrier IMF as in SDSS \citep[e.g.,][]{Bernardi10}. We also include in \figu\ref{fig|SigmaMbulge} the results of our Monte Carlo simulations with \modI\ in which we replace \mstar\ with \mbulge. It can be seen that the predicted, biased \sis-\mbulge\ relation from \modI\ (solid black lines) reproduces the measured slopes and normalizations in both samples (dashed blue lines).

\begin{figure*}
    \center{\includegraphics[width=15truecm]{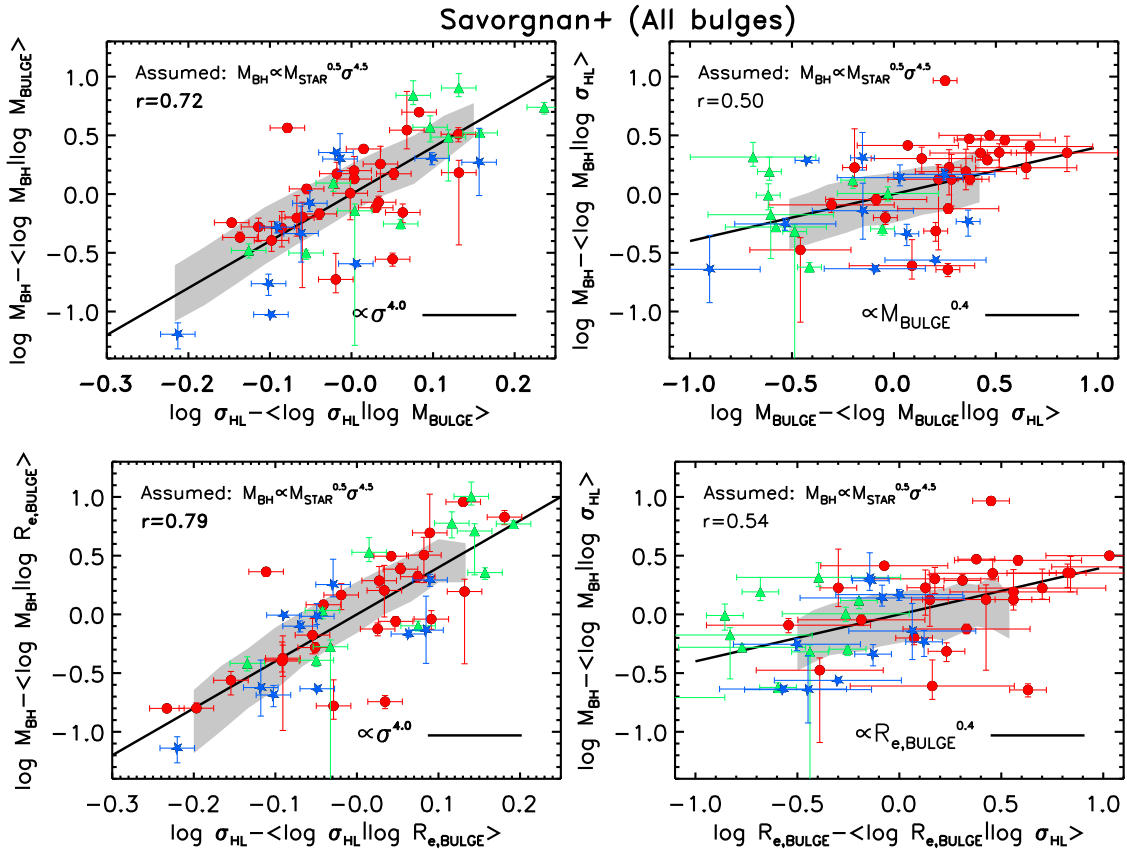}
    \caption{Same as \figu\ref{fig|MockResidualsModI} with data from the \citet{Savo15} sample of bulge stellar masses and effective radii. The red circles, green triangles, and blue stars refer to the bulges of ellipticals, S0s, and spirals, respectively. The grey bands show the predictions from our (selection biased) \modI\ with bulge stellar masses and effective radii. Spirals continue to show residuals similar to those of early-type galaxies, with the residuals in velocity dispersion still being much stronger than with other galactic properties, in agreement with \modI.
    \label{fig|MockResidualsMbulge}}}
\end{figure*}

\begin{figure*}
    \center{\includegraphics[width=15truecm]{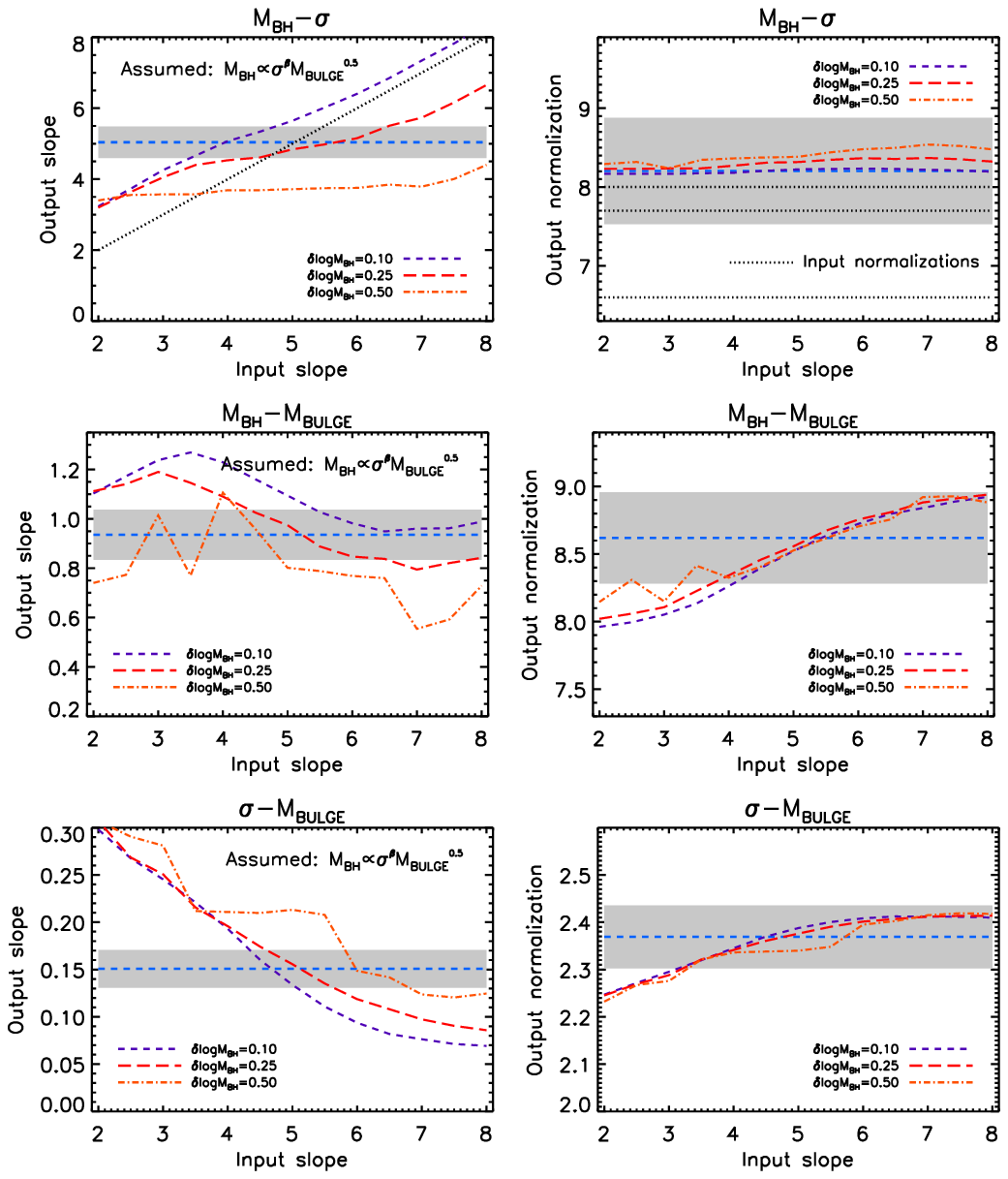}
    \caption{Dependence of the selection-biased slope (left) and normalization (right) on the intrinsic slope $\beta$ and scatter $\delta$ for our \modI\ ($\mbhe \propto \mstare^{0.5}\sigma^{\beta}$).  Upper, middle, and lower panels show results for the selection-biased $\mbhe-\sigma$, $\mbhe-\mstare$, and $\sise-\mstare$ relations.  Short-dashed, long-dashed and dot-dashed lines in each panel are for different values for the intrinsic scatter, as labelled. The dotted black lines in the right upper panel mark the three different normalizations chosen for each different value of the scatter; the lower normalizations correspond to higher values of $\delta$. The dotted black line in the upper left panel shows a simple one-to-one relation to guide the eye. The solid lines and grey regions mark the mean and dispersions in the E+S0 sample of \citet{Savo15}. Models with substantial scatter ($>0.3$ dex) and/or input slopes much flatter or steeper than $\beta\sim 5$ are disfavoured by the data.
    \label{fig|MultipleRuns}}}
\end{figure*}

In both panels of \figu\ref{fig|SigmaMbulge}, red circles identify disc galaxies with a bar.  These tend to have similar \sis\ to barless galaxies of the same \mbulge.  Thus, as we noted when discussing \figu\ref{fig|BiasedFJrels}, the selection bias does not seem to be affected strongly by the presence of a bar.

In view of the similarities between \figu\ref{fig|SigmaMstar} and \figu\ref{fig|BiasedFJrels}, it should not be to surprising that the corresponding \mbh-\sis\ and \mbh-\mbulge\ relations are also biased by the selection effect.  Therefore, rather than showing this explicitly, we consider the correlations between residuals from these scaling relations defined using bulge luminosities and effective radii.  \figu\ref{fig|MockResidualsMbulge} -- the analogue of \figu\ref{fig|MockResidualsModI} -- shows results for the \citet{Savo15} sample.  It is clear that spirals (blue stars) follow similar relations to those defined by earlier-type galaxies (red circles and green triangles).  The slopes, reported in Table~\ref{Table|Slopes} (column 4), match those from the E-S0 sample well (column 2), though the dependence on \sis\ is stronger and that on \mbulge\ weaker.  For completeness, column3, labeled ``All'', reports the results of using the total \mstar\ rather than \mbulge\ even for spirals.  The grey bands in the various panels show (selection biased) \modI\ when bulge stellar masses and effective radii are used (slopes reported in column 6 of Table~\ref{Table|Slopes}); these are in good agreement with the data.

In summary:  Bulges of spirals follow similar relations to those defined by earlier-type galaxies (red circles and green triangles).  They too show a stronger dependence on velocity dispersion (left panels) than other properties (right).  Moreover, the addition of spirals increases the baseline over which the relations on the left can be measured; this tightens the correlations with \sis\ and weakens those with the other properties. This justifies our earlier claim that our results are not much affected by the use of stellar or bulge mass, and also further supports the scenario in which spirals, or better their bulges, are correlated with their central black holes via a steep \mbh-\sis\ relation, similarly to early-type galaxies. Evidently, \sis\ is much more important than either total or bulge luminosity and/or size.

\subsection{Dependence on strength of correlation with \sis}
\label{subsec|MultipleRuns}

Comparison of our selection-biased Monte Carlo simulations with the observations suggests that \sis\ plays an important, if not fundamental, role in determining \mbh.  However, we have not yet explored the range of acceptable values of the free parameters in our \modI.  \figu\ref{fig|MultipleRuns} shows the result of setting $\mbhe \propto \mbulgee^{0.5}\sigma^{\beta}$ and producing Monte Carlo simulations for a range of values of the slope $\beta$, and rms scatter $\delta$ (blue short dashed, cyan long dashed, and red dot-dashed as labelled).  As mentioned in \sect\ref{subsec|simulations}, a larger $\delta$ results in a larger selection bias, so a lower input normalization is required to reproduce the same set of observational data.  Therefore, we reduce $\gamma$ when $\delta$ is large (black, dotted lines in the upper, right panel). In the specific, when $\delta=0.25$ dex we set $\gamma=7.7$, as in \eq\ref{eq|MbhSigma}, while $\gamma=8.0$ and $\gamma=6.6$ for $\delta=0.1,0.5$ dex, respectively.  Also, because of how our simulations are set-up, our $\delta$ includes a potential contribution from measurement errors on the value of \mbh.  So, while they may not be the truly intrinsic values, comparing them with data, to which measurement error has contributed, is meaningful.

The top panels show how the slope and zero-point of the selection-biased \mbh-\sis\ relation we measure in our Monte Carlo simulations depends on the intrinsic slope and scatter; the middle panels show a similar study of the selection biased \mbh-\mbulge\ relation; the bottom panels compare with the selection-biased \sis-\mbulge\ relation (left panel of \figu\ref{fig|SigmaMbulge}).  Dotted lines in the top two panels show the input values; these show that the bias increases -- slope decreases and zero-point increases -- as the rms scatter increases.  In all cases, the dashed, blue lines and grey bands show the range of slopes and zero-points which the bulge sample of \citet{Savo15} allow. Requiring the models to match these bands in all six panels shows that the intrinsic relation should have $\gamma \sim 7.7$, $\beta\sim 5$ and total scatter $\delta\sim 0.25$~dex.  While models with small values of the input (total) scatter, e.g. $\delta=0.1$, may also be acceptable, they tend to be less realistic in view of the non-negligible observational uncertainties in the dynamical \mbh\ estimates \citep[e.g.,][]{FerrareseFord}.  Allowing for 0.2~dex of \mbh\ measurement uncertainties, an observed scatter of $\delta\sim 0.25$ dex, allows for intrinsic scatter of $\sim 0.15$ dex.
Our results are in line with the Monte Carlo simulations by \citet{DaiMbhSigma} and \citet{Gultekin09} who both independently noticed that when the sphere of influence of the black hole is taken into account, the intrinsic slope in the \mbh-\sis\ relation becomes steeper $\beta\gtrsim 4-5$, and the normalization lower than what actually measured by a factor of $\gtrsim 2$. A more quantitative comparison with their results is hindered by the fact that their simulations differ substantially from ours, which are based on a realistic large distribution of local, unbiased galaxies with proper measurements of stellar masses, velocity dispersion and effective radii.

\sect\ref{sec|discu} discusses the implications of the values $\gamma \sim 7.7$, $\beta\sim 5$ and $\delta\lesssim 0.3$~dex for our understanding of the co-evolution of black holes and their host galaxies.

\subsection{Calibration of other \mbh\ proxies}
Without significant improvements in technology, it is difficult to measure dynamical masses at smaller \mbh\ locally, or at all at significant redshift.  The alternative is to look for other observational signatures of \mbh\ which do not require that $r_{\rm infl}$ be resolved.  Active galaxies are interesting in this respect since they can be observed both locally and at greater distances, so the same observational proxy for \mbh\ can be used over a wide range of redshift.  The key step in this process is to calibrate these proxies using the dynamical mass estimates we have been discussing so far.

Recently \citet{ReinesVolonteri15} have compiled a sample of 262 broad-line AGN at $z<0.055$ which roughly overlaps in volume with the dynamical mass \BH\ samples we have been discussing so far.  Their ``virial'' \mbh\ estimates depend on a constant of proportionality $f_{vir}\approx 4.3$, whose value was calibrated by matching to a \mbh-\sis\ relation which is like that of the selection biased sample of \citet{Savo15}.  With this calibration, they find that their AGN sample has $\log(\mbhe/M_\odot) = 7.45 + 1.05\log(\mstare/10^{11}M_\odot)$.  These \mbh\ values are substantially smaller than those for which dynamical \mbh\ measurements are available; the local (typically inactive) early-types have $\log\mbhe = 8.9 + 1.23\log(\mstare/10^{11}M_\odot)$.  Is physics responsible for this factor of $\sim 50$ discrepancy, or are selection effects also playing a role here?

Regarding selection effects in this context, \citet{Graham11} have argued that $f_{vir}\approx 2.8$ may be a more appropriate choice.  Our own analysis suggests that, because of selection effects, the \mbh\ values in the AGN sample should be reduced by a much larger factor, $\gtrsim 3$, making $f_{vir}\approx 1$.  In either case, reducing $f_{vir}$ would exacerbate rather than reduce the apparent difference with the local \mbh-\mstar\ relation.  However, this is not the full story.  Recall that the selection bias is much more dramatic for \mbh-\mstar\ than for \mbh-\sis\ (top panels of \figu\ref{fig|ScalingsMocks}).  So, it is possible that the AGN samples are probing lower masses where the \mbh-\mstar\ bias is particularly severe.

\begin{figure}
    \center{\includegraphics[width=8.4truecm]{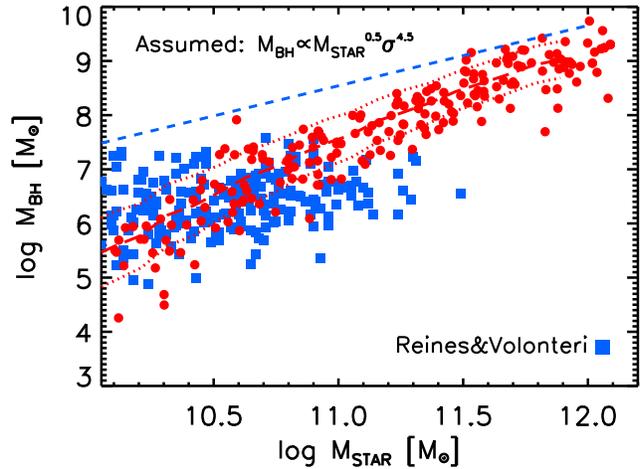}
    \caption{The \mbh-\mstar\ relation for the $z<0.055$ broad line AGN sample of \citet{ReinesVolonteri15} (blue squares), shifted downwards by a factor of $\sim 3$ as described in the text (i.e., $f_{vir}\approx 1$); the intrinsic relation for our \modI\ (red circles); and the corresponding selection biased relation in \modI\ (dashed blue line).  }
    \label{fig|AGN}}
\end{figure}

To illustrate, the blue squares in \figu\ref{fig|AGN} show the data of \citet{ReinesVolonteri15}, with their \mbh\ values lowered by the (factor of three on average) difference between the intrinsic and selection biased Monte Carlo samples in \modI\ (top left panel of \figu\ref{fig|ScalingsMocks}).  The blue dashed line -- which lies substantially above the AGN sample -- shows the selection biased \mbh-\mstar\ relation from the top right panel of \figu\ref{fig|ScalingsMocks}.
The corresponding intrinsic \mbh-\mstar\ relation is shown by the red circles.  The agreement between this intrinsic relation and the AGN sample is striking, particularly at $\log(\mstare/\msune) \sim 10.5$, where the bulk of the AGN data lie.

\begin{figure}
    \center{\includegraphics[width=8.4truecm]{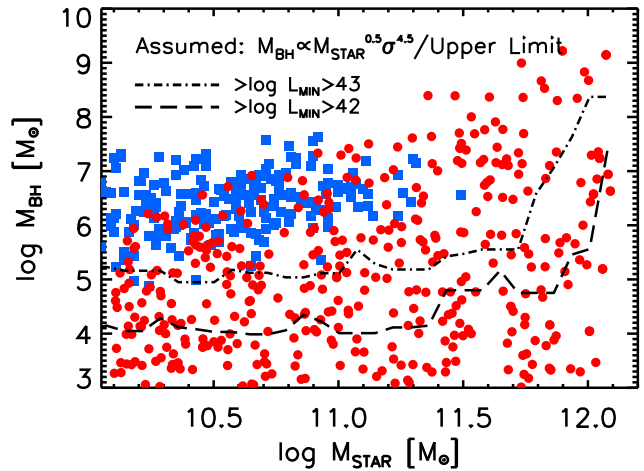}
    \caption{Predicted \mbh-\mstar\ relation for a model in which the scaling relation implied by \modI\ only represents an upper limit to the intrinsic distribution of black hole masses at fixed velocity dispersion (red circles). Dot-dashed and long-dashed lines show the smallest observable \mbh\ if the AGN is shining with the bolometric luminosity as labelled; the lower (long dashed) line is the luminosity cut imposed by \citet{ReinesVolonteri15}, whose data are shown by the blue squares.  The red circles which lie above this line show that, if the upper-limit model were correct, then many low mass black holes should have been detected.  Since there are no blue squares at these low masses, the data rule out this model.
    \label{fig|Batcheldor}}}
\end{figure}

Some of the remaining difference is a consequence of reporting results in terms of \mstar\ rather than \mbulge, as it is likely that the relevant comparison is with \mbulge\ \citep[see, e.g.,][]{GrahamScott15}.  Whereas \mstar$\approx$\mbulge\ for the early-types with dynamical \mbh\ measurements, \mbulge$<$\mstar\ for the AGN sample.  This would shift the blue squares to smaller \mstar\ (i.e. to the left in \figu\ref{fig|AGN}), further improving the match with the red circles (the unbiased \modI\ relation).  This agreement means that essentially all the offset between these AGN data and current local dynamical mass \BH\ samples is a selection effect, and that $f_{vir}\approx 1$.

More recently, \citet{Laesker16} analyzed nine megamaser disc galaxies with an average stellar mass of $\mstare \sim 10^{11}\, \msune$ and an average black hole mass of $\mbhe \sim 10^7\, \msune$.  They report an offset of $\delta \log \mbhe/\msune =-0.8\pm0.2$ with respect to the \citet{Laesker14} best-fit \mbh-\mstar\ relation for inactive galaxies, which is consistent with the mean difference between our observed and intrinsic relations in \modI\ (\figu\ref{fig|ScalingsMocks}). Similarly, \citet[][see also \citealt{HoKim16}]{HoKim14}, compared reverberation-mapped AGNs with measured bulge stellar velocity dispersions against the \mbh-\sis\ relation of inactive galaxies, finding a mean offset lower by $\delta \log \mbhe/\msune =-0.79$ dex (see their \figu2). The discussion above shows why we believe the discrepancy between the normalizations of the black hole scaling relations of active and dynamically-based samples is in large part a selection effect in the latter.

\subsection{On why we support the intrinsic black hole-galaxy scaling relation as a ridge}
\label{subsec|MbhSigmaRidge}

We complete this section by discussing the possibility put forward by \citet{Bat10} that the intrinsic black hole-galaxy scaling relation is not a relatively narrow ridge, but that the observed relation represents the upper limit of a much broader, almost uniform, distribution of $\log\mbhe$ at fixed \sis.  \citet{Ford98} had already argued that the lack of objects with small \mbh\ and large \sis\ means this is unlikely to be correct.  \citet{Gultekin11} added that if it were correct, there should be many more upper limits (i.e., non-detections, because $r_{\rm infl}\ll 0.1''$ for most objects) in the literature than detections -- but this is not observed.

To address this directly, we performed the same set of simulations as for \modI\ -- varying the input slope, normalization and scatter of the ``upper envelope'' \modI\ relation (\eq\ref{eq|MbhSigma}) -- and obtained results qualitatively similar to those reported in \figu\ref{fig|MultipleRuns}.
However, we believe that the relevant question is:  Does the intrinsic distribution of \mbh\ extend to much smaller values than our Monte Carlo simulations have assumed?

Since even local dynamical \mbh\ samples cannot probe small \mbh, we again turn to the AGN sample of \citet{ReinesVolonteri15}.  As \figu\ref{fig|AGN} shows, this sample clearly has many small \mbh\ at small \mstar\ but not at large \mstar.  (A rather tight correlation between low-mass active black holes and their hosts' spheroid stellar masses was also recently inferred by \citealt{GrahamScott15} by combining a number of independent data sets collected from the literature.)  The blue squares in \figu\ref{fig|Batcheldor} show this sample again.  The red circles show the expected distribution if \citet{Bat10} were correct; we assumed the intrinsic distribution was uniform in $\log\mbhe$ from an upper envelope defined by the observed \mbh-\sis\ relation down to $10^3\, \msune$, though our conclusions do not depend on the exact value.  This shows clearly that the upper envelope model is not consistent with the local AGN sample.  Had we not shifted the AGN samples downwards by a factor of three, the discrepancy with the red symbols would have been even more dramatic.

We have also carried out additional tests to probe the impact of flux limit effects on the observed distributions of \citet{ReinesVolonteri15}. We assigned broad, Schechter-type Eddington ratio distributions to our mock black holes in line with empirical estimates \citep[e.g.,][and references therein]{Shankar13acc,Schulze15}. Long-dashed and dot-dashed lines in \figu\ref{fig|Batcheldor} show the limiting active black hole mass that is still detectable above a bolometric luminosity of $L_{\rm bol} = 10^{42}\,\ergse$ and $L_{\rm bol} = 10^{43}\,\ergse$. It is clear that whatever the chosen input duty cycle of active black holes, at the minimum luminosity of $L_{\rm bol}=10^{42}\, \ergse$ probed by \citet{ReinesVolonteri15}, the detectable limit extends between one and two orders of magnitude below the data.  We conclude that a very broad distribution of local black holes down to very low masses, even at large \sis\ or \mstar, is not favoured by current data on local active galaxies (see further discussion in \sect\ref{sec|discu}).

\section{Discussion}
\label{sec|discu}

\subsection{Direct implications of the bias in the observed scaling relations between black holes and galaxies}
\label{subsec|discudirectimplications}

We have confirmed previous findings that all local galaxies with dynamical black hole mass estimates are a biased subset of all galaxies and this indicates that \BH\ scaling relations currently in the literature are biased \citep{Bernardi07}. Comparison of the selection-biased scaling relations in our Monte-Carlo simulations with observations, along with analysis of the residuals, strongly suggest that  the \mbh-\sis\ relation is fundamental, with a possible additional, relatively weak dependence on stellar (bulge) mass. In particular, our analysis suggests that the observed \mbh-\mstar\ relation, and as a consequence correlations with any other photometric property such as S\'{e}rsic index \citep[e.g.,][]{GrahamDriver}, are all highly biased. The intrinsic correlation between black hole mass and host galaxy stellar mass (total or bulge) is, according to our study, mostly a consequence of the \mbh-\sis\ relation. Including (the bulges of) spirals in our reference sample of ellipticals and lenticulars \citep[that of ][]{Savo15} confirms and extends our results. In this context it is no longer meaningful, at least within the biased samples, to look for outliers in the observed \mbh-\mstar\ relation, such as ``pseudo-bulges'' \citep[e.g.,][]{KormendyPseudo}, or examine whether bulge or total luminosity is a better predictor of \mbh\ \citep[e.g.,][]{Marconi03,HaringRix,Kormendy11b,Laesker14}, or consider the connection to nuclear star clusters only in terms of stellar mass \citep[e.g.,][]{Antonini15,Georg16}.

Our Monte Carlo results constrain the normalization of the intrinsic black hole-galaxy scaling relations to be a factor of $\gtrsim 3$ lower than current estimates, in terms of velocity dispersion, and up to a factor of $\sim 50-100$ lower when expressing black hole masses as a function of stellar mass (e.g., \figu\ref{fig|ScalingsMocks}). These results can reconcile the apparent mass discrepancies between local dynamical mass samples and local active galaxies (e.g., our \figu\ref{fig|AGN}), in Narrow Line Seyfert 1 active galaxies \citep[e.g.,][and references therein]{Orban11,Sani11,Mathur12,Shankar12,Calderone13}, moderately luminous AGN \citep[e.g.,][]{HoKim14,ReinesVolonteri15,HoKim16,Laesker16}, active low surface brightness galaxies \citep{Subra15}, and possibly also in more distant samples \citep[][]{San14,Falomo14,Busch15}.

The lowered normalization of the intrinsic \mbh-\sis\ relation will serve as a more secure base for calibrating virial estimators of black hole mass for reverberation mapping-based scaling relations \citep[e.g.,][]{Onken04,Woo10,Graham11,Park12,Grier13,HoKim15}. The exact value of $f_{vir}$, which depends on the structure, dynamics, and line-of-sight orientation of the broad line region, is indeed still a matter of intense debate \citep[see, e.g.,][for a recent discussion]{Yong16}. Typical values in the literature vary between average values of $f_{vir}\sim 1$ and $f_{vir}\sim 4$, depending, respectively, on whether the emission-line profile which enters the virial equation is measured via its full-width at half maximum or its line dispersion \citep[e.g.,][]{Collin06,Park12FWHM,HoKim14}. Our analysis tends to favour lower values of $f_{vir}$, i.e. $f_{vir}\approx 1$ rather than $4$. A smaller $f_{vir}$ by itself has a number of interesting consequences. For example, it may ease the challenge of growing very massive black holes in the early Universe \citep[e.g.,][]{Mortlock11,Trakh15,Wu15}, alleviates the need to invoke very massive seeds \citep[e.g.,][]{AleNat14,Madau14,Lupi15}, and it may also add some empirical evidence towards the existence of intermediate-mass black holes \citep[e.g.,][]{Farrell14}.

Having lower mass black holes may imply a proportionally lower integrated local black hole mass density \citep[e.g.][]{Tundo07,Bernardi07,GrahamDriverBHMF,YuLu08,SWM,Shankar13}, rather than a factor of a few higher as current estimates based on the (selection biased) \mbh-\mstar\ relation suggest. The most recent accretion models \citep{Shankar13acc,Aversa15}, based on Soltan-type \citep{Soltan} arguments and continuity equation models \citep[e.g.,][]{Cavaliere71,SmallBlandford,Marconi04,YuLu04}, suggest moderate average radiative efficiencies on the order of $\epsilon \lesssim 0.1$. Further increasing the local mass density by a factor of a few, as suggested by the current estimates of the local \mbh-\mbulge\ relation, would imply a radiative efficiency proportionally lower \citep{Novak13}, forcing the accretion models towards somewhat extreme scenarios such as frequent radiatively inefficient accretion and/or large fractions of heavily obscured, Compton-thick active galaxies \citep{Comastri15}. In contrast, a high radiative efficiency would imply that most of the local black holes are spinning rapidly, suggesting that spin may not be the only parameter controlling radio loudness in AGN, in line with many other, independent lines of evidence \citep[e.g.,][and references therein]{Sikora07,Shankar08Cav,Shankar10radio,Shankar16}.

Finally, lowering the normalization for the intrinsic black hole scaling relations impacts the expected signal in gravitational wave searches \citep[e.g.,][and references therein]{Sesana14,Sesana16}. A factor of few reduction in the normalisation reduces the characteristic strain amplitude arising from an incoherent ensemble of gravitational waves, and hence the expected signal-to-noise ratio of the gravitational wave background currently being searched for using pulsar timing arrays \citep[e.g.,][]{Sesana13b,Nano15,Rosado15,Sesana16,Simon16}.
While this reduction alleviates much of the tension between previous predictions and the lack of a detected signal \citep[e.g.,][]{Shannon15,Taylor16}, it also suggests that the detection of gravitational radiation via radio telescopes will be more difficult than previously thought \citep[e.g.,][]{Sesana16}.

\subsection{Implications for the co-evolution of black holes and galaxies}
\label{subsec|coevo}

In standard models, massive, bulged galaxies, the usual hosts of supermassive black holes, are formed in a highly star-forming, gas-rich phase at early cosmological epochs. A central, ``seed'' black hole is expected to gradually grow via gas accretion, eventually becoming massive enough to shine as a quasar and trigger powerful winds and/or jets that are capable of removing gas and quenching or inhibiting star formation in the host galaxy (``quasar-mode'' and ``radio-mode'' feedbacks). Feedback from an active black hole has indeed become a key ingredient in many galaxy evolution models \citep[e.g.,][]{Granato04,Bower06,Croton06,Monaco07,Sija15}. At later times, both the host galaxy and its black hole may further increase their mass (and size) via a sequence of mergers with other galaxies/black holes. Late mergers can contribute up to $\sim$80\% of the final mass \citep[e.g.,][]{DeLucia06,Malbon07,Marulli08,Naab09,ShankarBernardi09,Oser10,ShankarPhire,Gonzalez11,Shankar13,Shankar14,Shankar15,Zhang15,Zhao15}. The apparent tightness of the \mbh-\mstar\ relation is sometimes used to motivate black hole mass growth by dry mergers \citep[e.g.,][]{Peng07,JahnkeMaccio}.  Such arguments must be reconsidered if this tightness is just a selection effect (\figu\ref{fig|ScattersMocks}).  Whether merger models can explain the tightness of the \mbh-\sis\ relation remains to be seen.

Moreover, our results suggest that (the bulges of) spirals, which are usually not considered to have undergone a substantial phase of late dry mergers \citep[e.g.,][]{Huertas13b,Patel13b,Huertas15}, define similar correlations as do ellipticals, thus further pointing to the \mbh-\sis\ relation as the dominant correlation. This weakens the motivation for models in which \mbh\ in spirals grows substantially via any secular process unrelated to \sis\ \citep[e.g.,][]{Bower06,HopkinsHernquist09b,Bournaud11b,DraperBallantyne12,ShankarPseudo,FF15,GattiShankar}.

The importance of \sis\ inferred from our analysis supports models in which AGN, and in particular quasar-mode feedback, play a key role in linking black holes to their host galaxies. Our results suggest that the scaling with \sis\ is strong, $\mbhe\propto\sigma^5$ (\figu\ref{fig|MultipleRuns}), typical of energy-driven winds \citep[e.g.,][but see also \citealt{Cen15}]{SilkRees,Granato04,Hopkins06}.  In contrast, momentum-driven winds produce a somewhat weaker trend: $\mbhe\propto\sigma^4$, though the exact normalization and slopes predicted by AGN-feedback models as a function of time, mass, and host morphology is still a matter of intense debate \citep[e.g.,][]{Fabian99,King03,Fabian12,Fau12,Gabor14,King14}.

Evolution in scaling relations is a powerful constraint on models \citep[e.g.,][]{Merloni10}.  However, our results suggest that searches for evolution in terms of \mstar\ are less well-motivated (because the observed \mbh-\mbulge\ relation is more biased and less fundamental). On the other hand, most analyses based on Soltan-type arguments or direct detections suggest that the \mbh-\sis\ relation evolves weakly if at all \citep[e.g.,][but see also \citealt{Woo08}]{Gaskell09,ShankarMsigma,ZhangYuLu,Shen15}.  This further paves the way towards using black holes and quasars as cosmological distance estimators \citep[e.g.,][]{Hoenig14}.

Our results shed some light on the long-standing issue regarding the most accurate local black hole mass estimator in favour of the \mbh-\sis\ relation (with similar slope but lower normalization, and a possible additional weak dependence on stellar mass).  However, this raises a puzzle.  \citet{Lauer07demo} argued that the \mbh\--$L_{\rm bulge}$ relation is a more reliable \BH\ mass estimator for the most massive/luminous galaxies hosting the most massive black holes, such as those in the brightest cluster galaxies \citep[see also, e.g.,][]{Laporte14}.  They argued that \BH\ mass estimates via the \mbh\--\sis\ relation are too low to account for the existence of cores in these galaxies, believed to be produced by the ejection of stars during the decay of \BH\ binaries.  In this respect their conclusions are at odds with our argument that the \mbh-\sis\ relation is more fundamental than \mbh-\mstar.  However, our \figu\ref{fig|ScalingsMocks} shows that if \sis\ is the driving parameter, then the \mbh-\mbulge\ relation is at least as biased as the \mbh-\sis\ relation, so one should worry about how selection effects affect their argument.  In addition, it is possible that other processes such as dynamical friction and AGN feedback effects may contribute to the creation of cores in the inner regions of massive galaxies \citep[e.g.,][]{Elzant04,ToniniLapi,Martizzi13,Elzant16}.

A number of observations and black hole accretion models \citep[e.g.,][]{Marconi04,Merloni04,Granato06,Lapi06,Zheng09,Silverman08LF,SWM,Mulla12,Lapi14} suggest some degree of correlation between black hole growth and large-scale star formation. On the other hand, a number of observational and theoretical studies are now showing that the actual co-evolution may be more complex to probe observationally, possibly depending on the different evolutionary phases undergone by the host galaxies as well as AGN variability effects \citep[e.g.,][and references therein]{Hickox14,Rodi15,Volonteri15,GrahamReview15,West16}. Indeed the large scatter measured in the correlation between star formation and X-ray AGN luminosity \citep[][and references therein]{Dai15} might simply reflect the large scatter in the intrinsic \mbh-\mstar\ relation (\figu\ref{fig|ScattersMocks}), a possible independent sign for velocity dispersion, instead of stellar mass, acting as the main driver of the co-evolution between central black holes and their hosts.

How and why galaxies transition from a very active, star-forming phase to a red-and-dead one is still hotly debated. There are a number of hypotheses (not necessarily mutual exclusive) put forward in the literature to explain the quenching of star formation in galaxies \citep[e.g.,][]{Woo13}.  Our results on the importance of the \mbh-\sis\ relation suggests that the action of black hole feedback triggered during a quasar-like phase may be a substantial contributor to quenching.  In this respect, mounting evidence for a nearly environment-independent flattening in the star formation rate-stellar mass relation at high \mstar\ \citep[e.g.,][and references therein]{Erfa15}, paralleling an increased incidence of bulge-dominated galaxies, is also suggestive. The increase in host velocity dispersion may in fact be correlated with the growth of a central black hole; the associated feedback can reduce star formation, though alternative explanations in terms of, e.g., morphological transformations, may still be viable solutions \citep[e.g.,][]{Martig09,Huertas15}.

\section{Conclusions}
\label{sec|conclu}

The main aim of this work was to revisit the local scaling relations between \BHs\ and their host galaxies.  Our main results can be summarized as follows:
\begin{itemize}
   \item We have confirmed previous findings that local galaxies with dynamical black hole mass estimates are a biased subset of all galaxies \citep[e.g.,][]{YT02,Bernardi07,Remco15}.  At fixed stellar mass, local \BH\ hosts typically have velocity dispersions that are larger than the bulk of the population, irrespective of their exact morphological type or of the aperture within which the velocity dispersion aperture is measured (Figure~\ref{fig|SigmaMstar}). One of the main reasons for this bias is the observationally imposed requirement that the black hole sphere of influence (equation~\ref{eq|rinfl}) must be resolved for the black hole mass to be reliably estimated.
  \item We have confirmed the assertion in \citet{Bernardi07} that the selection bias cannot be ignored:  \BH\ scaling relations currently in the literature are biased.  To properly interpret the measured \mbh-scaling relations, one must quantify the effects of this bias.  We did so by carrying out Monte Carlo simulations in which we assumed different scaling relations to assign \BHs\ to mock galaxies (equations~\ref{eq|MbhSigma}--\ref{eq|MbhW}), and then applied the $r_{\rm infl}$-related selection cuts.  These show that the intrinsic relation
$$\log\left(\frac{\mbhe}{M_\odot}\right) = \gamma + \beta\,\log\left(\frac{\sigma}{200~{\rm km~s}^{-1}}\right)+ \alpha\, \log\left(\frac{\mstare}{10^{11}\, \msune}\right)$$
with $\gamma=7.7$, $\beta\sim 4.5-5$, $\alpha \lesssim 0.5$, and intrinsic scatter of the order of 0.25~dex reproduces all the observed scaling relations (\figus\ref{fig|ScalingsMocks}, \ref{fig|MockResidualsModI} and \ref{fig|MultipleRuns}), as well as the biased relation between velocity dispersion and stellar mass observed in local \BH\ hosts (\figu\ref{fig|BiasedFJrels}).  Equations~(\ref{eq|IntrinsicMbhMstar}) and~(\ref{eq|IntrinsicMbhSigma}) give the \mbh-\mstar\ and \mbh-\sis\ relations which result from this \mbh-\sis-\mstar\ relation.
\item The observed \mbh-\mstar\ relation is much more biased than \mbh-\sis\ (\figu\ref{fig|ScalingsMocks}).  The apparent tightness of the \mbh-\mstar\ relation is a selection effect, as are trends of the scatter with mass (\figu\ref{fig|ScattersMocks}).
 \item A more detailed comparison of the scaling relations in our selection-biased Monte Carlo samples with similar relations in real data suggest that the correlation with velocity dispersion is the dominant one:  any additional dependence on stellar mass and effective radius must be small (\figus\ref{fig|MockResidualsModI}, \ref{fig|MockResidualsModII} and~\ref{fig|Appendix2}--\ref{fig|Appendix5}).
 \item Spirals tend to define similar correlations to ellipticals and lenticulars.  All our results remain valid if we replace \mstar\ with \mbulge\ (Section 4.5; \figus\ref{fig|SigmaMbulge}, \ref{fig|MockResidualsMbulge} and~\ref{fig|Appendix1}).
\end{itemize}

\noindent Our findings have a number of implications:\\
(1) Our preference for steeper slopes in the intrinsic \mbh-\sis\ relation ($\beta \gtrsim 5$), is consistent with that of energy-driven AGN feedback models.  Our normalization, $\log(\mbhe/\msune)=7.8$ at $\sigma=200\, \rm{km\, s^{-1}}$ (\eq\ref{eq|IntrinsicMbhSigma}), is a factor $\gtrsim 3\times$ lower than previous estimates.  This suggests proportionally lower black hole mass densities, and so higher radiative efficiencies, supporting a scenario in which most super-massive black holes are rapidly spinning.  Reducing \mbh\ values by a factor $\gtrsim 3$ also reduces the predicted gravitational wave signal from black hole mergers, perhaps explaining why pulsar timing arrays have not yet reported detections.  \\
(2) Our revised intrinsic black hole scaling relations will serve as a more secure base for calibrating virial estimators of black hole mass for reverberation mapping-based scaling relations.  Our results suggest that the calibration factor should be reduced to $f_{vir}\approx 1$.\\
(3) The fact that the \mbh-\mstar\ relation is so much more biased than \mbh-\sis\ (top panels of \figu\ref{fig|ScalingsMocks}) explains most of the offset between local (inactive) \mbh\ samples having dynamical mass estimates, and AGN-based samples (\figu\ref{fig|AGN}).  \\
(4) Our simulations also disfavour broad distributions of black hole masses at fixed velocity dispersion (\figu\ref{fig|Batcheldor}). \\
(5) Unless one has accounted for selection effects, looking for outliers (e.g. bars, pseudo-bulges) from the \mbh-\mstar\ relation is no longer so meaningful.  Similarly, searches for redshift evolution in the \mbh-\mstar\ relation, which do not account for selection effects, are not well-motivated.  Since the apparent tightness and mass dependence of the \mbh-\mstar\ relation are biased by the selection effect, any heavily (dry) merger-driven black hole growth model must be reconsidered.\\
(6) The similarity of spirals to ellipticals means that the motivation for models which trigger black holes via disc instabilities or processes other than quasar feedback that do not directly involve velocity dispersion should be re-evaluated.\\
(7) As \sis\ is the controlling parameter in a number of other galaxy scaling relations \citep{Bernardi05}, our finding that \sis\ is the most important parameter in \mbh\ scaling relations will serve as a more robust test for the next generation of galaxy-black hole co-evolution models, for a deeper understanding of high-redshift data on active and star-forming galaxies, and for more accurate estimates of the black hole mass function.

\section*{Acknowledgments}
FS warmly thanks S. Ottavi for her invaluable and constant support. We thank A. Beifiori for sharing her data in electronic format and for helpful clarifications on some of the samples considered in this work. We thank an anonymous referee for helpful suggestions that improved the presentation of our results. FS also thanks N. McConnell, P. Monaco, A. Reines, and M. Volonteri for useful discussions.

\footnotesize{
  \bibliographystyle{mn2e_Daly}
  \bibliography{../RefMajor_Rossella}
}

\appendix

\textbf{\section{Residuals in other data sets}}
\label{sec|Appendix1}

\begin{figure*}
    \center{\includegraphics[width=14truecm]{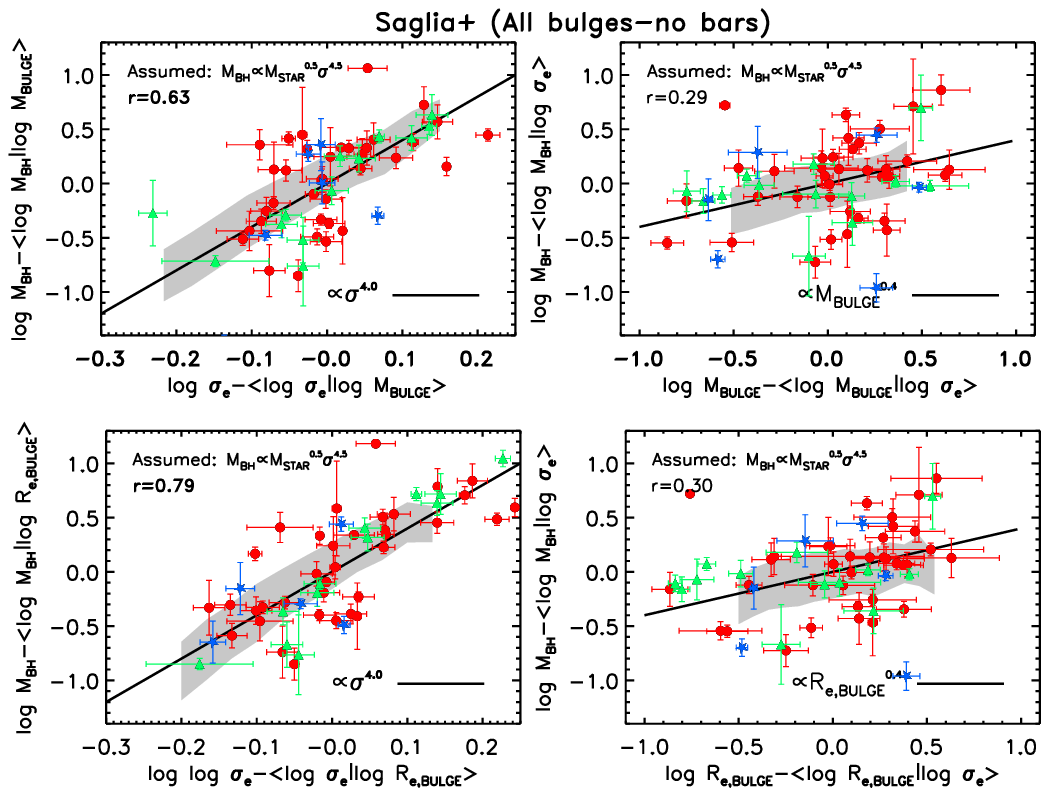}
    \caption{Same as \figu\ref{fig|MockResidualsMbulge}, but for bulge stellar masses and half-light radii from \citet{Saglia16}.  Red circles indicate barred galaxies, while the colour coding is otherwise the same as in \figu\ref{fig|MockResidualsMbulge}.
    \label{fig|Appendix1}}}
\end{figure*}

\begin{figure*}
    \center{\includegraphics[width=14truecm]{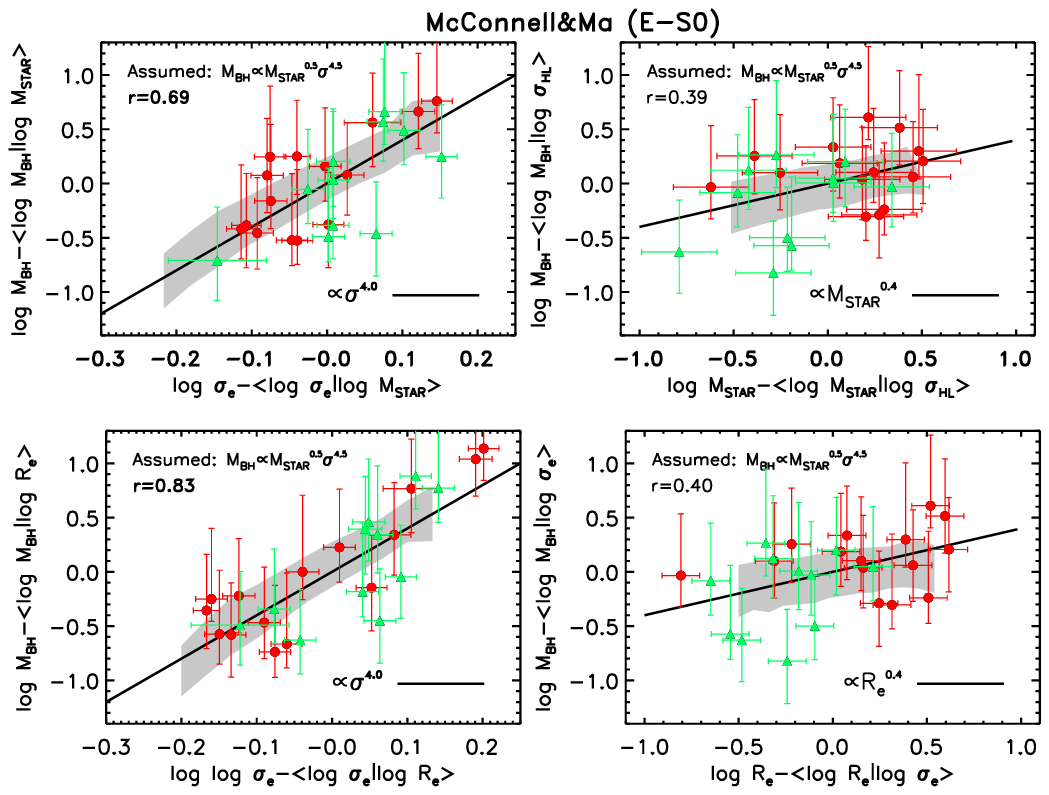}
    \caption{Same as \figu\ref{fig|MockResidualsModI} but using the E+S0 sample from \citet{Laesker14}. The data are consistent with velocity dispersion being the most fundamental property connecting black holes to galaxies.
    \label{fig|Appendix2}}}
\end{figure*}

\begin{figure*}
    \center{\includegraphics[width=14truecm]{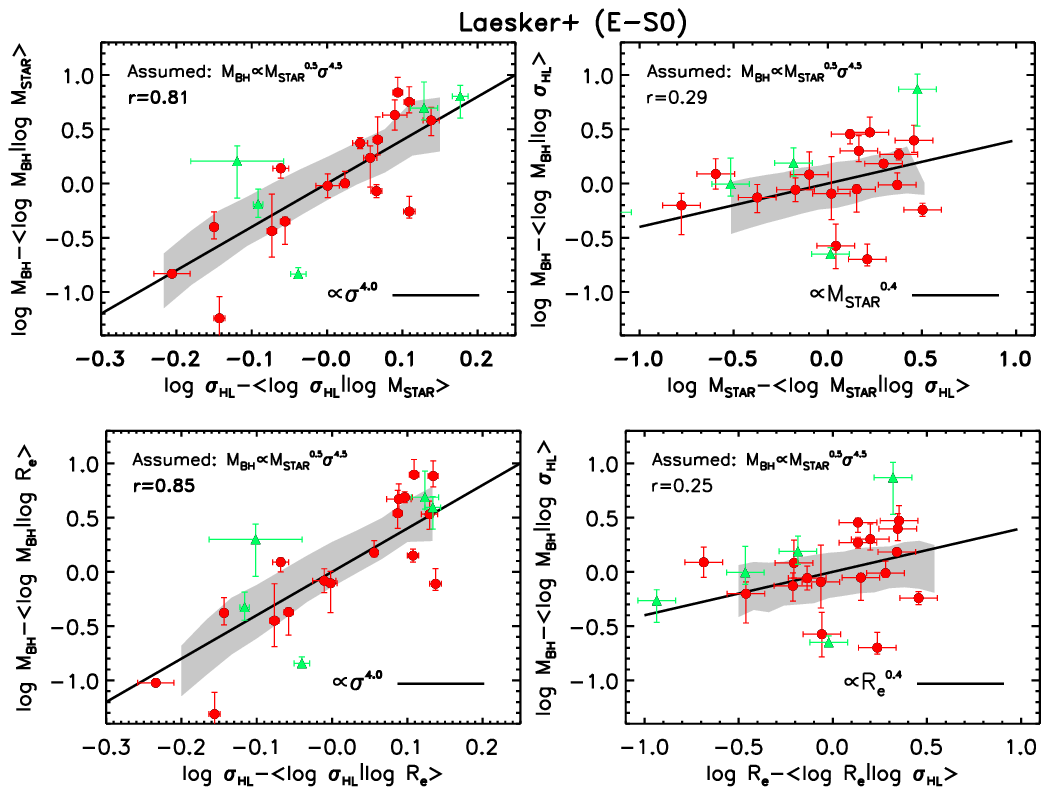}
    \caption{Same as \figu\ref{fig|MockResidualsModI} but using the E+S0 sample from \citet{Laesker14}. The data are consistent with velocity dispersion being the most fundamental property connecting black holes to galaxies.
    \label{fig|Appendix3}}}
\end{figure*}

\begin{figure*}
    \center{\includegraphics[width=14truecm]{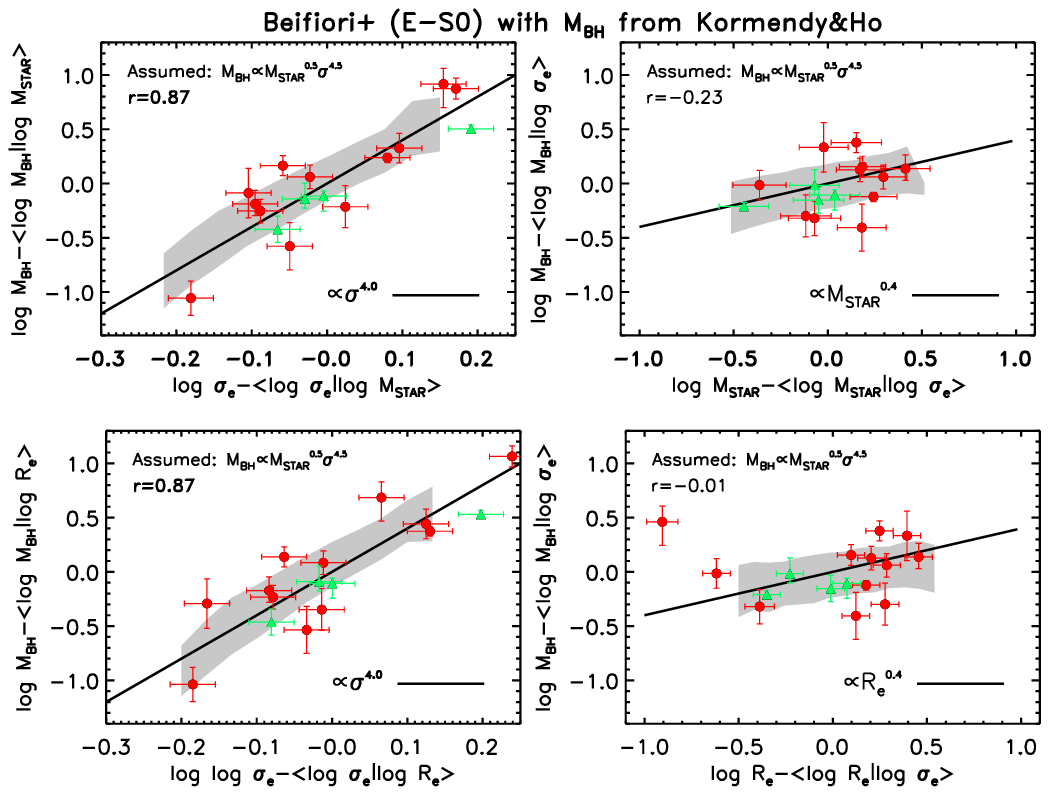}
    \caption{Same as \figu\ref{fig|MockResidualsModI} but using the E+S0 sample from \citet{McConn13} with galaxy $3.6\mu$ galaxy luminosities and effective radii from \citet{Sani11}. The data continue being consistent with velocity dispersion being the most fundamental property connecting black holes to galaxies.
    \label{fig|Appendix4}}}
\end{figure*}

Figures~\ref{fig|Appendix1}, \ref{fig|Appendix2}, \ref{fig|Appendix3}, and~\ref{fig|Appendix4} show correlations between the residuals from scaling relations measured in the samples of \citet{Saglia16}, \citet{McConn13}, \citet{Laesker14}, and \citet{Beifiori12} (the latter with black hole masses taken from \citet{KormendyHo}). \figu\ref{fig|Appendix1} shows the residuals using bulge luminosities and effective radii in the \citet{Saglia16} sample, excluding non-barred spirals, while all other samples refer to total stellar masses and effective radii of E+S0s (bulge properties are not available for most of these samples).  The grey bands show the corresponding correlations for \modI. In the panels on the left, all data sets define comparable if not even tighter correlations than those shown in \figu\ref{fig|MockResidualsModI}, while those on the right show weaker dependence on any other variable, in excellent agreement with \modI.

\textbf{\section{Residuals in \modIII}}
\label{sec|Appendix2}

\begin{figure*}
    \center{\includegraphics[width=15truecm]{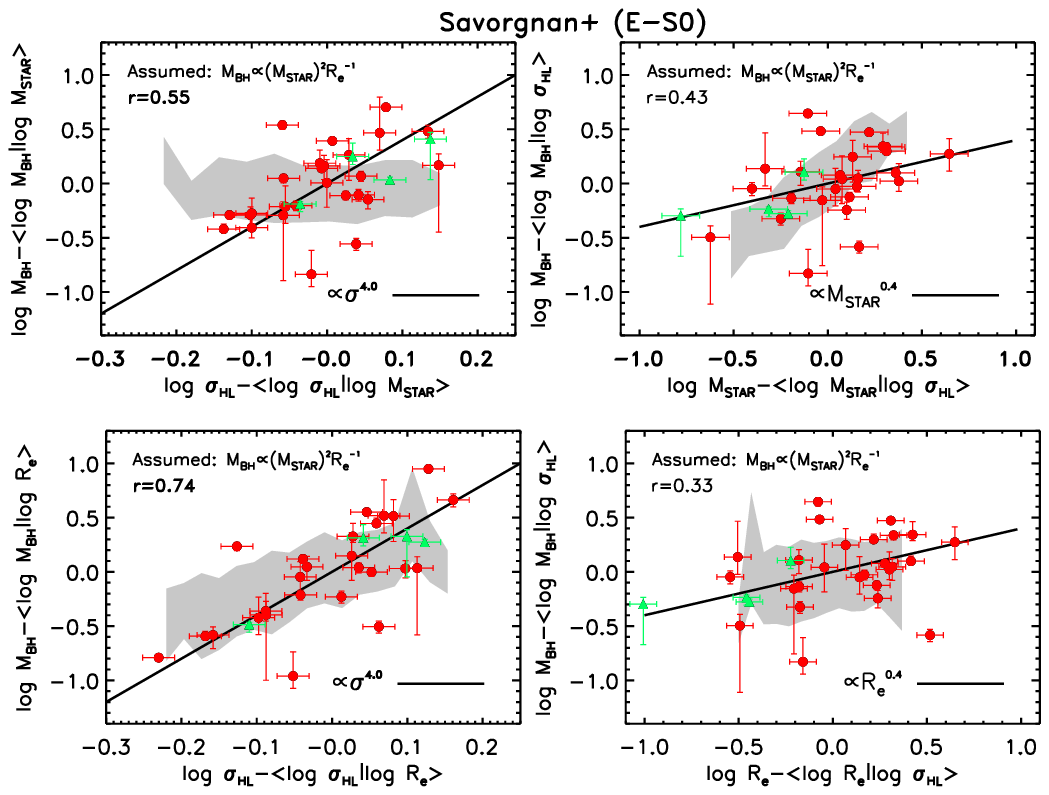}
    \caption{Same as \figu\ref{fig|MockResidualsModI} but for \modIII. Similarly to \modII, this model also predicts weaker correlations with $\sigma$ and stronger trends with other variables than is observed.
    \label{fig|Appendix5}}}
\end{figure*}

In \modIII, the intrinsic relation is $\mbhe\propto \mstare^2/R_e$, so residuals from correlations with velocity dispersion should be uncorrelated.  The grey band in the top left panel of \figu\ref{fig|Appendix5} shows this is also true in the selection biased sample.  In contrast, the \citet{Savo15} data show a strong correlation.  The data shows a steeper correlation in the bottom left panel as well.  On the other hand, in the top right panel, it is the model which  shows a stronger correlation than the data.  The slopes of these correlations, with uncertainties derived from many Monte Carlo realizations (see \sect\ref{subsec|MockResiduals}), are reported in Table~\ref{Table|Slopes}.

\label{lastpage}
\end{document}